\documentclass[letterpaper,twocolumn,10pt]{article}
\usepackage{usenix2019_v3}

\usepackage{geometry}
\newcommand{\fakepara}[1]{\vspace{1mm}\noindent\textbf{#1}\hspace{1.5mm}}

\newcommand{\eg}{\textit{e.g.}\xspace}
\newcommand{\ie}{\textit{i.e.}\xspace}
\newcommand{\cf}{cf.\xspace}

\newcommand{\lessthan}{\raisebox{0.07\baselineskip}{\relsize{-1}{\textless}}\hspace{0.3mm}}
\newcommand{\greaterthan}{\raisebox{0.07\baselineskip}{\relsize{-1}{\textgreater}}\hspace{0.3mm}}

\newcommand{\circone}{\protect\raisebox{-0.5pt}{\ding{192}}\xspace}
\newcommand{\circtwo}{\protect\raisebox{-0.5pt}{\ding{193}}\xspace}
\newcommand{\circthree}{\protect\raisebox{-0.5pt}{\ding{194}}\xspace}
\newcommand{\circfour}{\protect\raisebox{-0.5pt}{\ding{195}}\xspace}
\newcommand{\circfive}{\protect\raisebox{-0.5pt}{\ding{196}}\xspace}
\newcommand{\circsix}{\protect\raisebox{-0.5pt}{\ding{197}}\xspace}
\newcommand{\circseven}{\protect\raisebox{-0.5pt}{\ding{198}}\xspace}

\usepackage[utf8]{inputenc} %
\usepackage[table]{xcolor}
\usepackage{xspace} %
\usepackage{cite} %
\usepackage{paralist} %
\usepackage{graphicx} %
\usepackage{tikz} %
\usepackage{subcaption} %

\usepackage[framemethod=tikz]{mdframed} %

\usepackage[super]{nth}

\usepackage{bbm} %
\usepackage{hyperref}
\usepackage{tabu} %
\usepackage{longtable}
\usepackage{booktabs}
\usepackage{makecell} %
\usepackage{arydshln} %
\usepackage[usestackEOL]{stackengine} %
\usepackage{enumitem} %
\usepackage{pifont} %
\usepackage{stmaryrd} %
\usepackage{relsize} %
\usepackage[most]{tcolorbox} %
\tcbuselibrary{breakable}
\usepackage{adjustbox} %
\usepackage{array} %
\usepackage{multirow} %
\usepackage{verbatimbox} %
\usepackage{caption} %
\usepackage{layouts} %
\usepackage{bm} %
\usepackage[hang,flushmargin,bottom]{footmisc} %
\usepackage{scalerel}
\usepackage{soul}
\usepackage{fancyvrb}

\usepackage{tabulary} %

\usepackage{csquotes}
\usepackage[vskip=1pt]{quoting}
\SetBlockEnvironment{quoting}

\usepackage{epigraph} %
\usepackage{calc}
\newcommand{\mytextformat}{\rmshape\epigraphsize}

\let\originalepigraph\epigraph 
\renewcommand\epigraph[2]%
  {\setlength{\epigraphwidth}{\widthof{\mytextformat#2}}\originalepigraph{#1}{#2}}

\usepackage{tikz} %

\usepackage{titlesec} %
\titlespacing*{\section}{0pt}{2mm}{1mm}  %
\titlespacing*{\section}{0pt}{2mm}{1mm}  %
\titlespacing*{\subsection}{0pt}{2mm}{1mm}  %
\titlespacing*{\subsubsection}{0pt}{2mm}{0.5mm}  %

\usepackage{setspace}  %
\usepackage{marginnote}
\usepackage{tabularx}

\usepackage{url}
\PassOptionsToPackage{colorlinks}{hyperref}
\PassOptionsToPackage{pdftex}{hyperref}

\usepackage[nameinlink,noabbrev]{cleveref} %
\DeclareCaptionLabelSeparator{forcespace}{~} %

\definecolor{figurecolor}{RGB}{22,90,220}
\definecolor{citecolor}{RGB}{198,81,19}
\usepackage{url}
\makeatletter
\g@addto@macro{\UrlBreaks}{\UrlOrds}
\makeatother

\def\Snospace~{\S{}}

\usepackage{inconsolata}
\newcommand{\fun}[1]{{\texttt{\textls[-50]{#1}}}}
\newcommand{\funn}[1]{\hspace{-0.2mm}\textls[-50]{\texttt{#1}\xspace}\xspace}

\newcommand{\sampled}{\funn{sampled}}
\newcommand{\traceid}{\funn{traceId}}
\newcommand{\traceids}{\funn{traceIds}}
\newcommand{\bufferid}{\funn{bufferId}}
\newcommand{\bufferids}{\funn{bufferIds}}
\newcommand{\triggerid}{\funn{triggerId}}

\newcommand{\hsbegin}{\funn{begin}}
\newcommand{\hsend}{\funn{end}}
\newcommand{\tracepoint}{\funn{tracepoint}}

\newcommand{\trigger}{{trigger}\xspace}
\newcommand{\triggers}{{triggers}\xspace}
\newcommand{\triggered}{{trigger}ed\xspace}
\newcommand{\autotrigger}{autotrigger\xspace}
\newcommand{\autotriggers}{autotriggers\xspace}

\newcommand{\ucA}{\textcolor{blue}{UC1}\xspace}
\newcommand{\ucB}{\textcolor{blue}{UC2}\xspace}
\newcommand{\ucC}{\textcolor{blue}{UC3}\xspace}

\hypersetup{ colorlinks=true, urlcolor=black, linkcolor=figurecolor, citecolor=citecolor, pdfstartview=FitH
        pdftitle={The Benefit of Hindsight: Tracing Edge-Cases in Distributed Systems},
        pdfauthor={Lei Zhang, Vaastav Anand, Zhiqiang Xie, Ymir Vigfusson, Jonathan Mace}
        }

\definecolor{summarycolor}{rgb}{0.57, 0.36, 0.51}

\usepackage{listings}

\definecolor{codegreen}{rgb}{0,0.6,0}
\definecolor{codegray}{rgb}{0.5,0.5,0.5}
\definecolor{codepurple}{rgb}{0.58,0,0.82}
\definecolor{backcolour}{rgb}{0.96,0.96,0.96}
\lstdefinestyle{mystyle}{
    backgroundcolor=\color{backcolour},   
    commentstyle=\color{codegreen},
    keywordstyle=\color{magenta},
    numberstyle=\tiny\color{codegray},
    stringstyle=\color{codepurple},
    basicstyle=\ttfamily\footnotesize,
    breakatwhitespace=false,         
    breaklines=true,                 
    captionpos=b,                    
    keepspaces=true,                 
    numbers=left,                    
    numbersep=5pt,                  
    showspaces=false,                
    showstringspaces=false,
    showtabs=false,                  
    tabsize=2
}
\usepackage[ruled,vlined]{algorithm2e}

\begin{document}

\title{The Benefit of Hindsight: Tracing Edge-Cases in Distributed Systems}

\author{
\begin{tabular}{c c}
{\rm Lei Zhang} %
& \qquad
{\rm Vaastav Anand}\\[-0.5mm]
Emory University
& \qquad
Max Planck Institute for Software Systems\\[8pt]
{\rm Zhiqiang Xie} 
& \qquad
{\rm Ymir Vigfusson}\\[-0.5mm]
Max Planck Institute for Software Systems
& \qquad
Emory University\\[8pt]
\multicolumn{2}{c}{\rm Jonathan Mace} \\[-0.5mm]
\multicolumn{2}{c}{Max Planck Institute for Software Systems}
\end{tabular}
}

\maketitle

\begin{abstract}

Today's distributed tracing frameworks are ill-equipped to troubleshoot rare edge-case requests.
The crux of the problem is a trade-off between specificity and overhead.
On the one hand, frameworks can indiscriminately select requests to trace when they enter the system (head sampling), but this is unlikely to capture a relevant edge-case trace because the framework cannot know which requests will be problematic until after-the-fact.
On the other hand, frameworks can trace everything and later keep only the interesting edge-case traces (tail sampling), but this has high overheads on the traced application and enormous data ingestion costs.

In this paper we circumvent this trade-off for any edge-case with symptoms that can be programmatically detected, such as high tail latency, errors, and bottlenecked queues.
We propose a lightweight and always-on distributed tracing system, Hindsight, which implements a \emph{retroactive sampling} abstraction: instead of eagerly ingesting and processing traces, Hindsight lazily retrieves trace data only \emph{after} symptoms of a problem are detected.
Hindsight is analogous to a car dash-cam that, upon detecting a sudden jolt in momentum, persists the last hour of footage. 
Developers using Hindsight receive the exact edge-case traces they desire without undue overhead or dependence on luck.
Our evaluation shows that Hindsight scales to millions of requests per second,
adds nanosecond-level overhead to generate trace data, handles GB/s of data per node, transparently integrates with existing distributed tracing systems, and
successfully persists full, detailed traces in real-world use cases when edge-case problems are detected.

\end{abstract}

\section{Introduction}
\label{sec:intro}

Troubleshooting failures and performance problems in  large-scale distributed systems is crucial. On one side, tiny performance misbehavior in a production system could be costly~\cite{awsoutage, businesslosingindowntime, 40801}. On the other side, exacerbated by growing system complexity, diagnosing problems takes onerous effort from system developers and requires significant engineering resources.
Distributed tracing is invented as the solution of troubleshooting distributed systems by recording detailed,
end-to-end traces of requests executions,
and are proved helpful for a wide range of use cases~\cite{shkuro2019mastering,parker2020distributed}.

\newpage

Prior distributed tracing works have demonstrated a wide range of use cases. Common-case analysis focuses on aggregated system behaviors, such as monitoring resource usage~\cite{sambasivan2016principled, parker2020distributed, shkuro2019mastering,mann2011modeling,thereska2006stardust}. In contrast, edge-case troubleshooting (\autoref{sec:case_studies}), the topic of this paper, focuses on rare and outlier system behavior, such as tail latency~\cite{dean2013tail,li2014tales,zhang2016treadmill,suo2016time,misra2019managing}. %

Since an edge case is rare by definition, tracing edge cases requires trace coverage of all requests.
In typical production environments, tracing \emph{every} request---including transmitting, processing, and storing comprehensive telemetry---requires enormous backend infrastructure and storage that is unacceptable to infrastructure operators. 
State-of-the-art tracing frameworks manage this overhead by
collecting a small sample (0.001\%) of traces~\cite{las2019sifter, sigelman2010dapper, kaldor2017canopy, jaeger1861}.
Though previous works practically reduce tracing overhead through head sampling~\cite{sigelman2010dapper, kaldor2017canopy} and tail sampling~\cite{las2019sifter, lascasas2018weighted} techniques, they cannot trace edge cases at scale (\autoref{sec:edge_case_sampling}).

In this paper, we resolve the problem of tracing edge-case requests in production environments.
To achieve this, we focus our attention on \emph{symptomatic} edge cases, where the performance effects 
of the problem manifest shortly after its causes and where the impacts can be observed programmatically.
We propose \textbf{retroactive sampling}
to collect telemetry data \emph{back in time} from the present moment of detection from all machines that serviced the request.
The key idea is to generate all trace data but only collect useful data through a retrieval mechanism.

To implement retroactive sampling, we built Hindsight---an always-on, lightweight distributed tracing system %
that is compatible with existing tracing APIs---as a practical tool for edge-case analysis.
Under retroactive sampling, all trace data is recorded locally but only reported when a symptom is detected, allowing applications to generate copious trace data in case they are needed without
encumbering the tracing system's backend collection infrastructure. 
Retroactive sampling ultimately reports the same volume of 
trace data as other sampling methods, but ensures that edge-case traces are not missed.
To provide efficient and coherent retroactive sampling, Hindsight's design separates its dataplane, \eg generating trace data into fast local memory, from control logic, \eg for indexing metadata, coordinating among machines, and triggering collection for symptomatic requests on demand.

\newpage

As demonstration, we apply Hindsight on three use cases corresponding to our running examples. We run experiments on the DeathStar Microservices Benchmark~\cite{gan2019open}, the Hadoop Distributed File System~\cite{hdfs}, an Alibaba benchmark derived from production traces~\cite{luo2021characterizing}, and on several micro-benchmarks.  We have integrated Hindsight with OpenTelemetry~\cite{opentelemetry} and as a replacement collection component for X-Trace~\cite{fonseca2007xtrace}.  Our experimental results show that Hindsight imposes nanosecond-level overhead when generating trace data, can scale to GB/s of data per node, rapidly reconstructs traces when triggered, and effectively captures problematic traces, as well as related lateral 
traces, within tens of milliseconds of identifying a symptom.

In summary, our paper makes the following contributions.
\begin{compactitem}
\item We describe the retroactive sampling abstraction for capturing traces of symptomatic edge-cases.
\item
We present the design of Hindsight, a distributed tracing system that implements retroactive sampling. Hindsight is compatible with existing tracing APIs and can be transparently integrated with existing applications.
\item
We apply Hindsight on real-world use cases and show that efficiently collecting edge-case requests is practical.
\item 
We evaluate Hindsight on multiple benchmarks and real systems, showing that it can achieve nanosecond-level overhead on trace data generation and handle GB/s data per node while collecting coherent traces.
\item
We illustrate  that Hindsight is compatible and performs better than state-of-the-art tracing systems (X-Trace and Jaeger) with more efficient trace-data generation and lower overhead, while providing edge-case tracing. 
\end{compactitem}

\section{Motivation}
\label{sec:motivation}

\subsection{Edge-Case Troubleshooting}
\label{sec:case_studies}
\label{sec:motivation:edge_case_troubleshooting}

Consider the following three examples of real-world use cases \ucA--\ucC of edge-case troubleshooting from prior work.

\fakepara{Error diagnosis (\ucA).}
Hardware failures, component errors, exceptions, and programming mistakes abound in large production systems~\cite{yin2011empirical}. %
Application developers often play the role of detective, to identify root causes of errors.
An error might only arise due to a specific, rare combination of factors and code paths exercised;
the symptoms of a problem often manifest far from the root causes~\cite{mace2015pivot, erlingsson2012fay, luo2018troubleshooting}, and the potential root causes are manifold, perhaps combined  software or hardware problems on many nodes or network links~\cite{kannan2021debugging}.

\fakepara{Tail-latency troubleshooting (\ucB).}
Distributed systems track a wide range of high-level health metrics, such as API distributions, latency percentiles, resource utilization, and many others~\cite{kaldor2017canopy, srebook}.  
An operator may observe an unusual metric jump, say the $99^\text{th}$ percentile latency has spiked for some important API. 
However, knowing about the spike is not enough; the application developer must identify the specific service, code paths, or conditions that contribute to the peak to address any underlying problems~\cite{li2014tales,dean2013tail,misra2019managing}.

\fakepara{Temporal provenance (\ucC).}
Many modern distributed systems respond to requests through an architecture of loosely coupled microservices~\cite{shkuro2019mastering}.
Application developers need tools for tracking queuing issues when the number of components in a distributed system is large~\cite{hbase8228, hdfs6110,hbase8744, hdfs3751, hdfs11461}, since a request $R$ exhibiting symptoms (\eg prolonged queueing time) may not be the true culprit for the backlogged queue. Rather, the developer
wants to follow the \emph{temporal provenance} of $R$ to determine \emph{lateral traces} of other related requests with which $R$ interacted through shared components and queues~\cite{wu2019zeno}.

\subsection{Distributed Tracing} %
\label{sec:motivation:distributed_tracing}

Distributed tracing frameworks are in widespread use in both open-source~\cite{opentelemetry, zipkin, jaeger} and major internet companies~\cite{sigelman2010dapper, kaldor2017canopy, netflixTracing} to chronicle \emph{end-to-end requests}. 
A trace is a recording of one request, and each trace contains spans, events, and annotations, along with timing and ordering, generated from every machine visited by the request.
Compared to  traditional logs and metrics, the key distinction of distributed tracing is that a trace captures the full end-to-end structural flow of request execution across all components visited.

\fakepara{Advantages.}
Distributed tracing is thus particularly useful for troubleshooting cross-component problems in large systems, since the request traces explicitly tie together the individual slices of work performed across different machines, enabling an operator to observe how the work done by one machine influences, and is influenced by, work done on others~\cite{fonseca2007xtrace, sigelman2010dapper, netflixTracing, sridharan2018distributed}.
Prior research on distributed tracing demonstrates a range of use cases, including common-case analyses centered on aggregate system behavior, distributions over data, and relationships between system components~\cite{sambasivan2016principled, parker2020distributed, shkuro2019mastering, kaldor2017canopy}.

\fakepara{Limitations.}
Since edge-case troubleshooting concerns rare and outlier system behavior, the symptoms and evidence of a problem might only manifest in a very small fraction of requests. Unfortunately for the operator, this sparsity may yield exceptionally few exemplar traces of edge-case behaviors and symptoms, owing to the design of modern distributed tracing frameworks.
Let us look closer at how traces are captured before returning to this problem.

\begin{figure}[t]%
\centering%
\includegraphics[width=\columnwidth]{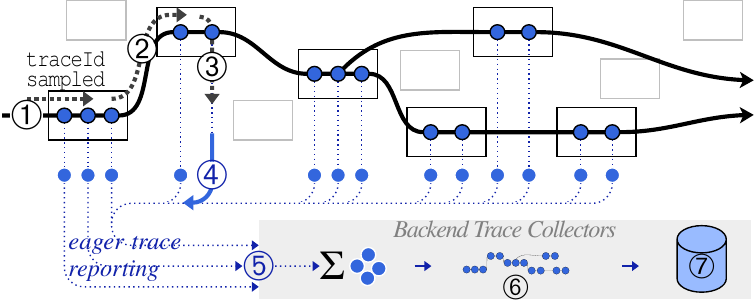}%
\caption{\textbf{Distributed tracing}~(\autoref{sec:motivation:distributed_tracing}).  A request (solid black line) traverses system processes, depositing trace data that is eagerly ingested into the trace collector backends.  End-to-end trace objects are constructed from trace data, processed, and stored in a database.
}%
\vspace{-0.2in}
\label{fig:tracing_overview}%
\end{figure}

\fakepara{Current designs.}
\autoref{fig:tracing_overview} depicts a typical distributed tracing framework~\cite{kaldor2017canopy, netflixTracing, opentelemetry}.
When a new request arrives at the application, the tracing framework assigns it a unique \traceid (\circone).  Every request is assigned a \traceid, but not every request is actually traced; the framework sets a per-request \sampled flag to indicate as such.  From this starting point, the application then propagates the \traceid and \sampled flag alongside the request at the application level and includes them with all inter-process communication (\circtwo).  

Any component that handles the request can generate trace data (\eg spans, events) using the tracing framework's client library (\eg OpenTelemetry~\cite{opentelemetry})---trace data is only generated if \sampled is set.  Trace data gets explicitly annotated with the \traceid, thereby associating the data with the request (\circthree).  Ultimately there may be many components and machines that handled the request and contribute trace data.  At the same time, many requests may execute concurrently (\eg in different server handler threads), generating temporally-interleaved data with different \traceids.

The framework's client library eagerly enqueues, serializes, and transmits trace data (\circfour) to its centralized backend collection infrastructure, or \emph{backend} for short~(\circfive). The backend is distinct from the traced applications and is responsible for continually receiving, processing~(\circsix), and storing~(\circseven) trace data generated across all of the application's components.  The backend uses the \traceid to \emph{join} data that was dispersed across many machines but belongs to same request into a single coherent trace object. The backend finally persists that trace object in a database if it decides to retain the trace.

\fakepara{Overhead vs.~incompleteness.}
Traces can be detailed and produced at high volume, risking overheads.  Traces at Google, for instance, are typically more detailed than debug-level logging~\cite{sigelman2010dapper}; each traced request at Facebook, similarly, generates several MBs of tracing data and approximately 1 billion traces are captured per day~\cite{kaldor2017canopy}. 
At high rates, tracing frameworks may encounter several potential bottlenecks: when generating data within the traced application (\circthree); when transmitting trace data over the network (\circfour); and in backend processing and storage (\circfive--\circseven).

To reduce overheads, the de facto practice is to capture fewer traces. 
Here, operating at the granularity of an entire trace maintains \emph{trace coherence}: %
if a request is sampled, then the whole trace is kept including all data across all machines; otherwise nothing is kept.  Coherent traces are essential for distributed tracing -- a partial or fragmented trace has limited value in diagnosis~\cite{fonseca2007xtrace, sigelman2010dapper, honeycombSampling, kamonIncoherentEdgecases} because it loses the end-to-end visibility that makes the trace useful in the first place~\cite{netflixTracing, sridharan2018distributed, parker2020distributed}.
There are two main approaches for foregoing traces coherently:
the system may decide to omit a request at \circone before tracing and ingestion (\emph{head sampling}) or the traces may be filtered after collection at \circsix  (\emph{tail sampling}).

\emph{Head sampling} reduces overheads by simply tracing fewer requests in the first place, \ie by setting the \sampled flag for only a small fraction of requests (\circone).  By leaving \sampled unset for the majority of requests, trace data will not be recorded for most requests, thus avoiding application overheads to generate data, ingestion overheads to transmit and process data, and storage overheads (\circthree--\circseven).  Head sampling is widely used in practice; it is enabled by default in Jaeger~\cite{jaeger} with a 0.1\% sampling probability, and some production systems sample as few as 0.001\%~\cite{sigelman2010dapper, kaldor2017canopy}.

\emph{Tail sampling} is used to drop traces at the trace backends (\circsix).  Unlike head sampling, the application will still trace all requests and will incur all expenses of generating and ingesting the trace data (\circthree--\circfive).  Tail-based sampling primarily allows backends to lower the trace storage costs by selectively dropping traces after combining them into trace objects but before committing them to storage~\cite{las2019sifter, lascasas2018weighted, otTailSampling}.

\subsection{Edge-Case Troubleshooting Troubles}
\label{sec:edge_case_sampling}
 
Recall that edge-case problem symptoms only manifest in a small fraction of requests, which are undetermined until the problem takes place. We argue that current approaches are ineffectual at getting traces of edge-cases.

\fakepara{Head sampling sacrifices edge-cases.}  
Indiscriminate sampling decisions made at the \emph{beginning} of a request (\circone), while useful for
curbing overhead, \emph{cannot} know a priori whether a request will encounter a rare edge-case problem and should be traced. 
For edge-case troubleshooting this presents an obstacle: a low head-sampling probability (\eg 0.1\%) means a trace of the problem will exist with low probability (\ie 0.1\%). 
The developer may thus have reports that errors took place (\ucA) yet the corresponding `rare' requests were not sampled when those requests began---they lack the detailed cross-machine data necessary for finding the error's root cause.
Likewise, the application's high-level metric monitoring may indicate a spike in end-to-end tail-latency (\ucB); the developer is thus aware that these high-latency outliers exist, yet without a trace, they cannot localize the problem to a particular component or request class.
The situation is even more problematic when investigating bottlenecked queues via temporal provenance (\ucC): since each request was sampled independently, the tracing system will have only a vanishing probability that traces of \emph{all} relevant requests in the queue were captured.

\fakepara{Tail sampling sacrifices overheads and scalability.}
Practitioners have long pointed out a discord between what traces are interesting and what traces get head-sampled~\cite{zipkinSecondarySampling, jaeger1861, oteps115, ots307,jaeger425}.  Fortunately, many common edge-case symptoms, including error codes (\ucA) and high end-to-end response time (\ucB), can be recorded directly within the trace data itself.  This enables tail-samplers to explicitly seek out edge-case traces, because at this point (\circsix) they can directly inspect the constructed trace object.  Today's tail-samplers support filtering traces based on span attributes or metrics, thereby targeting a range of outlier symptoms such as high tail latency, unexpected error codes, uncommon attributes, rare code paths, and undesirable behavior such as RPC retries~\cite{jaeger425, grafanaHorizontalScalability, splunkUseCases, newrelicTailsamplingUsecases, otTailSampling, newstackTracing, netflixTracing, ot407}.

Tail-sampling entails enormous costs, however: they must trace all requests and ingest all trace data in order to make informed decisions.  Application latency and throughput can suffer if tracing libraries lack optimization (\eg 2$\times$ throughput reduction using OpenTelemetry tail-sampling~(\autoref{sec:eval:application_overhead})).
Ingesting all traces consumes substantial network bandwidth between applications and collectors, interfering with latency-sensitive application traffic (\eg up to 200\,MB/s per node~(\autoref{sec:eval:application_overhead})). Tail-sampling demands large backend infrastructure investment, deploying enough collectors to receive and process all incoming traces (\eg even one chatty RPC server can overwhelm an OpenTelemetry collector (\autoref{sec:eval:application_overhead})).  
Even assuming perfect horizontal scaling, tail-sampling requires \eg $100\times$ the collector capacity of 1\% head-sampling.
Lastly, tracing frameworks are also not robust to bottlenecks mid-way through ingestion (\eg network backpressure) and quickly lose trace coherence when overloaded.

\fakepara{Practitioners sacrifice edge-cases.}
The justified pragmatism of avoiding large overheads means that head sampling reigns supreme in real-world distributed tracing deployments~\cite{kamonIncoherentEdgecases, honeycombWhitepaper, elasticco, kaldor2017canopy, sigelman2010dapper}.
Even tail-sampling features of commercial products have low thresholds on data ingestion (\eg \lessthan350\,kB/s per host~\cite{splunkTraceLimits}, \lessthan34\,spans/s per host~\cite{newrelicTailbasedLimits}, \lessthan6\,MB/s per collector~\cite{lightstepMicrosatellites}) after which vendors will automatically enable head-sampling or incoherently drop spans.
Ultimately, the operator who wishes to troubleshoot edge cases is left unfulfilled.

\section{Approach}
\label{sec:approach}

Hindsight aims to overcome today's trade-off between overheads and edge cases.  Our goal is to enable practitioners to target edge-case traces with the flexible criteria of tail sampling, while retaining overheads similar to that of head-sampling, \ie without high application overheads or substantial additional backend infrastructure. We now describe several insights that lead us to Hindsight's \emph{retroactive sampling} approach.

\fakepara{It is not expensive to generate trace data.}
We don't know a priori whether a request will be an interesting edge-case; only after symptoms manifest.  Paradoxically once we observe symptoms, it is too late to just enable tracing from that point on, as we have already missed the events that led to the anomaly.  The only sure-fire way of obtaining coherent traces for any edge-case is to record trace data from the very beginning of the request, for every request.

Tail-sampling does just that -- with high overheads and steep infrastructure costs.  However, this is primarily because today's tracing frameworks tightly couple trace generation with trace ingestion.  Ingesting data is expensive, incurring network and backend infrastructure costs.  Generating data into local memory is not -- outside of distributed tracing, \eg, we observe new technology like Intel PT can generate 100--200\,MB/s of processor telemetry per core at 5--15\% runtime overhead~\cite{intelpt}; likewise method-tracing techniques for Android applications exhaustively record all function entries and exits with \lessthan1\,ns per tracepoint and \lessthan3\% runtime overhead~\cite{luo2022hubble}.  We believe that comparable overheads should be possible for distributed tracing.  With careful client library design, applications should be able to generate detailed trace data locally into memory, in anticipation of that data being useful if a problem occurs.

\newcommand{\cemph}[1]{\noindent\begin{mdframed}[hidealllines=true,backgroundcolor=blue!5,    
innertopmargin=3pt,
innerbottommargin=3pt,
innerrightmargin=2pt,
innerleftmargin=2pt,
leftmargin = 0pt,
rightmargin = 0pt,
skipabove = -12pt,
skipbelow = -12pt
]{#1}%
\end{mdframed}\vspace{-0.5em}}

\vspace{-0.1in}
\cemph{\emph{Retroactive sampling:} nodes generate, but do not ingest, all trace data.}
\vspace{-0.05in}

\fakepara{Symptoms are locally observable.}  Although root causes are many, varied, and difficult to predict, the same is not true of \emph{symptoms} of problems.  For example, error codes, tail latency, and exceptions are easily-observed indicators of potential problems.  Many symptoms are localized, programmatically detectable, and manifest quickly at some point during or shortly after a request was served~\cite{huang2018capturing, netflixTracing, otTailSampling}.  For example, tail sampling techniques, by definition, require that some span in the trace was explicitly annotated with the symptom of an anomaly, and typically wait only 10 seconds to accumulate trace data~\cite{newrelicTailbased, otTailSampling}.  For these common cases it is not necessary to ingest and construct full trace objects when the symptom is so readily detectable at the source.  Moreover, since symptoms can be detected independent of traces in the first place,  we do not need the expensive indirection of writing symptoms into trace data only to later extract and filter them.  We believe that the key to capturing edge-cases is to decouple detection of symptoms from collection of traces.

\vspace{-0.1in}
\cemph{\emph{Retroactive sampling:} applications embed \textbf{triggers} that programmatically observe symptoms and signal after-the-fact that a trace is an edge-case.}
\vspace{-0.05in}

\fakepara{Triggers are local but trace data is distributed.}  Prior distributed tracing frameworks ingest traces eagerly.  We instead believe that traces should be lazily ingested, only in response to a trigger fired at some point during or soon after a request.  However, triggers are local -- only one machine might detect a symptom, yet the trace data for the request will be dispersed across memory of all machines that serviced it. To splice together a coherent end-to-end trace, all of these other machines need to learn of the trigger and send their slice of the trace to the backend collectors. To identify and notify all relevant machines of a trigger, we thus need the ability to \emph{back-track} the end-to-end path of a request.

\vspace{-0.1in}
\cemph{\emph{Retroactive sampling:} requests propagate and deposit \textbf{breadcrumbs} so triggers can be shared with all relevant machines.}
\vspace{-0.05in}

\fakepara{Trace data will eventually expire.}  Applications generate trace data into local memory where it incurs no further processing.  We only send trace data to collector backends if a trigger fires.  However, we cannot predict \emph{when} a trigger might fire -- even if a request has finished executing locally, we cannot easily know that the request isn't still executing on some other machine(s) or that a trigger won't fire remotely.  Thus, trace data must remain in memory on each machine indefinitely.  Over time this will fill memory and eventually we will need to free up space.  The intuitive choice is thus to expire trace data for the least-recently-seen request.  We call the implicit time duration between generating data and overwriting it the \emph{\textbf{event horizon}}.  We believe that retroactive sampling should not require a large event horizon -- as low as tens of seconds is reasonable -- because triggers are automatic and shared quickly.  In the majority of cases a machine should learn of a trigger within a matter of seconds or milliseconds.  Thus retroactive sampling should be feasible even with large and detailed traces or constrained memory.

\vspace{-0.1in}
\cemph{\emph{Retroactive sampling:} triggers are best effort; we assume we will see triggers quickly if at all.}
\vspace{-0.05in}

\section{Design} %
\label{sec:design}

\label{sec:compatibility}
\fakepara{Overview.}
Hindsight is a distributed tracing framework that implements retroactive sampling.
Whereas typical distributed tracing frameworks eagerly ingest trace data, Hindsight lazily ingests data only after a trigger, thus allowing retroactive sampling of edge-case traces without paying the overhead costs of ingesting all trace data.
Hindsight remains compatible with existing head-sampling and tail-sampling policies.  Hindsight trivially implements head-sampling policies by firing an immediate \trigger upon a positive head-sampling decision (or if the \sampled flag is set).  Hindsight is opaque to backend trace collectors and tail-sampling policies, and existing ingestion pipelines require no changes.  Likewise, Hindsight is transparently compatible with
existing OpenTelemetry APIs and instrumentation~\cite{opentelemetry}, and piggybacks breadcrumbs with OpenTelemetry's context propagation.

\begin{figure}%
\centering%
\includegraphics[trim=0 3 6 0,clip,width=\columnwidth]{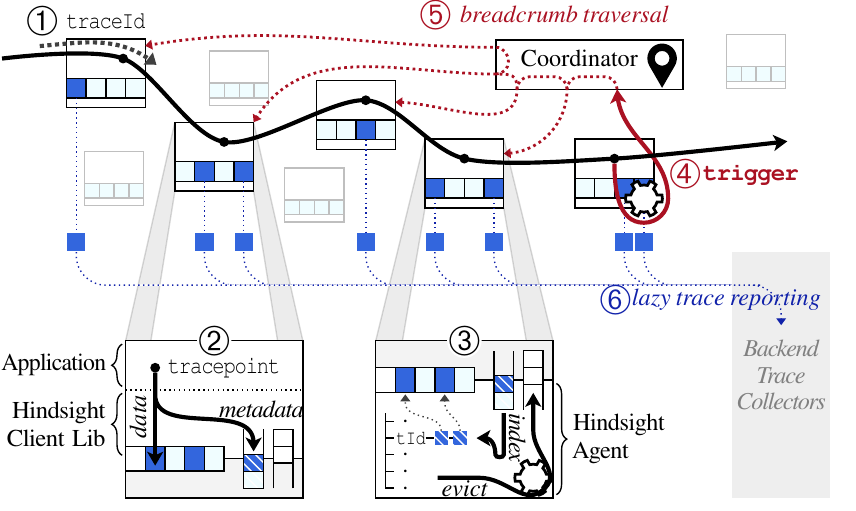}%
\caption{\textbf{The end-to-end lifecycle of a trace in Hindsight}~(\autoref{sec:hindsight:overview}).%
}%
\vspace{-0.1in}%
\label{fig:arch}%
\end{figure}

\fakepara{Walkthrough.}
\label{sec:hindsight:overview}
\autoref{fig:arch} shows a high-level diagram of Hindsight's main components.
\noindent\begin{itemize}[noitemsep,topsep=0pt,parsep=0pt,partopsep=0pt,label={},leftmargin=0.5cm]
\item[\hspace{0mm}\circone] On request arrival (solid black line) Hindsight generates a unique \traceid and thereafter propagates it alongside the request, as done by existing frameworks (\autoref{sec:motivation:distributed_tracing}).
\item[\hspace{0mm}\circtwo] Applications record trace data (\eg events, spans) using Hindsight's \tracepoint client API.  This leaves the request's trace data scattered across the machines it visited.
\item[\hspace{0mm}\circthree] A Hindsight \emph{\textbf{agent}} runs on each machine to manage trace data.  Hindsight agents do not inspect, process, or eagerly report trace data to backends -- instead, agents index metadata by \traceid and await further instruction.  For most traces nothing further happens, the trace is not reported, and agents eventually evict old trace data.
\item[\hspace{0mm}\circfour] If an application node observes an outlier symptom (\eg erroneous response, high latency, or a bottlenecked queue) it invokes Hindsight's \trigger API and passes the request's \traceid.  
\item[\hspace{0mm}\circfive] The local Hindsight agent receives the triggered \traceid.  The full trace remains dispersed across many Hindsight agents, so the local agent informs Hindsight's logically centralized \emph{coordinator} service of the \traceid.  Hindsight's coordinator recursively contacts the set of machines that serviced this request, soliciting breadcrumbs deposited by the request at each machine; a breadcrumb is a pointer to another machine involved in the request (\eg to the RPC caller or callee).
\item[\hspace{0mm}\circsix] Each agent contacted will set aside its slice of data belonging to the \traceid, and asynchronously send it to the backend collector.%
\end{itemize}

\fakepara{Design decisions.}
Hindsight is most shaped by three key design choices.
First, to prioritize trace coherence as a primary objective throughout the architecture. 
Second, to maintain an efficient data and control plane split to enable tracing 100\% of requests.
Finally, to support lightweight programmatic trigger mechanisms.

\subsection{Trace Coherence}
\label{sec:design:coherence}

Coherence is a top-level requirement for distributed tracing (\autoref{sec:motivation:distributed_tracing}).  As soon as any machine drops data for a trace, the trace is incoherent and effectively useless for troubleshooting.  Hindsight's design  avoids incoherence in several places.

At \circthree, agents continually evict old trace data to free up space for new data.  Agents do this atomically at the granularity of a trace; there is no point in only dropping part of a trace.  However, for a single trace, its data is non-contiguous and fragmented in memory.  Agents carefully organize and index metadata about where each trace's data resides and do not simply evict old data in a LIFO manner.

At \circfive, the coordinator must contact all agents that handled a request before those agents overwrite their slice of trace data.  Breadcrumbs are a lightweight and scalable solution -- the coordinator recursively follows breadcrumbs and only contacts the specific agents known to have serviced the request.  This approach takes only a few milliseconds in our evaluation.  Breadcrumb traversal is independent of reporting the trace data; agents set aside and asynchronously send trace data to the collector backends after learning of a trigger.

At \circsix, agents can potentially experience network congestion or backpressure from the collector backends, such as in response to a trigger-happy application that fires too many triggers and causes a backlog of unreported trace data on many, or all, agents.  Eventually even triggered data must be dropped.  Hindsight agents do not drop data arbitrarily (\eg, skipping a full queue) because different agents would then tarnish different victim traces---it only takes one agent dropping its slice of a trace to render the remaining data on other agents practically worthless due to incoherence.  Instead, in several places agents use priority queues, with priority determined by consistent hashing of \traceids.  A given \traceid will enjoy the same priority across all agents and queues, and the same traces will be dropped by all agents in the face of a bottleneck.

Finally, at \circfour, applications may fire multiple different \triggers for a diverse range of symptoms, using a developer-provided \triggerid to distinguish different trigger types.    Hindsight will prevent a profuse trigger from stifling trace collection of other, low-frequency triggers: agents implement weighted fair sharing for reporting and evicting trace data, with user-defined  weights and rate-limits for each \triggerid.

\subsection{Efficient Data Management}
\label{sec:design:datamanagement}
Lazy ingestion significantly reduces the volume of trace data sent from agents to the backend trace collection infrastructure.  However, within an individual machine, retroactive sampling requires the application generate trace data into local memory for \emph{all} requests (\circthree).  The most sensitive performance bottleneck for Hindsight is thus between client applications generating data (\tracepoint) and the local Hindsight agent that manages trace metadata.  
Our design establishes a clear split between \emph{control} and \emph{data} activities, which congregates general-purpose data and efficiency in the data plane, and embeds all logic in the control plane.

\fakepara{Data plane.}
Hindsight's {data plane} is concerned with efficiently writing trace data from client applications.  Using \tracepoint, applications write trace data to a large shared memory pool subdivided into \emph{buffers}.  Different threads write to different buffers; each buffer may only belong to one \traceid at a time, and threads acquire new buffers when full or when the active \traceid changes.
Consequently, the buffer pool is not consumed sequentially and a single trace may be fragmented across several non-contiguous buffers.

\fakepara{Control plane.}
Hindsight's agent process encapsulates {control plane} activities,
continually circulating \emph{metadata} about buffers to the application, via two shared memory queues.  Applications poll for available buffers and push full buffers; agents poll for full buffers, index metadata of full buffers grouped by \traceid, and push evicted buffers back to the application.  Agents receive triggers and communicate with Hindsight's coordinator, manage breadcrumbs linking the trace data that is strewn across many agents, extract triggered trace data, and report data asynchronously to the backend trace collection infrastructure.  Hindsight's control and data distinction yields an efficient agent implementation because agents only touch metadata.

\subsection{Triggers}
\label{sec:design:triggers}

Applications initiate retroactive sampling via Hindsight's \trigger API (\circfive).
In the common case, symptoms are easy to detect and localize: top-level error codes; high latency; increased queue time.
Such symptoms can be readily recognized and cheaply computed without the trigger mechanism needing the trace data itself.
For example, this may entail adding a \trigger call within a service's exception handler, or after checking for outlier latency upon a request's completion. 
Hindsight provides a library of automatic triggers based on metric percentiles, categorical features, and exceptions. All of our use cases (\ucA--\ucC) can be implemented using Hindsight's \autotriggers.  Likewise all existing tail-sampling policies can be implemented using \autotriggers, as span-local attribute and metric filters directly translate to metric and categorical \autotriggers.  

By separating triggers from traces, developers can also implement custom symptom detectors to explicitly decide the conditions for triggering.  It further leads to a straightforward integration of triggers into existing metric-monitoring and outlier-detection systems regardless of their architecture.

\fakepara{Lateral traces.}
Outlier behavior may not map directly to a single request; instead there may be several other related \emph{lateral} requests.
For example, to diagnose a bottlenecked queue (\ucC), a trigger needs to capture traces for the previous $N$ requests to understand what led to queue buildup~\cite{wu2019zeno}; to diagnose a write-ahead log, we desire all requests blocking on a log sync~\cite{hbase8228, hdfs6110}; to diagnose resource contention we require all requests contending for a slow disk or network~\cite{hbase8744, hdfs3751, hdfs11461}.
By separating triggers from traces, we enable more comprehensive trigger conditions based on factors beyond just a single trace, and triggers that can capture multiple related traces simultaneously.
Hindsight enables an application to atomically trigger a group of related lateral \traceids; internally Hindsight will ensure that the group as a whole is coherently collected.  By comparison, tail sampling cannot easily express cross-trace triggers or sample lateral traces, because \traceid-based sharding in collector backends is fundamentally at odds with sharing state between traces.

\section{Implementation}
\label{sec:implementation}

We have implemented Hindsight's client library in $\approx$4KLOC of C and Hindsight's agent and coordinator in $\approx$5.5KLOC of Go.
We chose C for dataplane efficiency and Go for its ease of use for the more complex control plane logic.

\subsection{Data Plane Buffer Pool}

Each Hindsight agent pre-allocates a fixed-size \emph{buffer pool} in shared memory for applications to directly write trace data.  
Hindsight logically subdivides the buffer pool into fixed-size buffers (default 32\,kB).
Client applications write trace data to buffers via Hindsight's client API.  The agent process does not touch data in the buffer pool except when reporting \triggered traces.
At each point in time, a buffer can only contain trace data of a single request; no two different requests will write trace data to the same buffer at the same time.  
A single trace will thereby comprise (1) multiple non-contiguous buffers on each agent and (2) many buffers scattered across numerous agents.
Buffers are the granularity of data management within Hindsight.  Within clients and agents, a buffer is addressable by its \bufferid---its offset into the buffer pool.

\definecolor{antiquefuchsia}{rgb}{0.57, 0.36, 0.51}
\begin{table}[t]%
\rowcolors{1}{blue!3}{blue!9}%
\noindent%
\centering%
\relsize{-0.5}{%
\resizebox{\columnwidth}{!}{%
\centering%
\begin{tabular}{p{0.4\columnwidth} p{0.6\columnwidth}}
\fun{begin(traceId)} &  Request begins in the current thread. \\
\fun{tracepoint(\{payload\})} &  Record data for the current trace; \fun{payload} is of arbitrary size in bytes. \\
\fun{breadcrumb(address)} &  Adds a breadcrumb to the current trace, pointing to some other node \fun{address}. \\
\fun{serialize()} &  Obtain the current \traceid and a \fun{breadcrumb} to the current node. \\
\fun{end()} &  Request ends processing in current thread; flush and remove buffers.\\
\fun{trigger(traceId,triggerId,} \fun{lateralTraceIds...)} & Instruct Hindsight to collect \traceid and zero or more \fun{lateralTraceIds}%
\end{tabular}%
}}%
\vspace{-2mm}
\caption{\textbf{Hindsight client API.} Applications can invoke the API directly, or indirectly using Hindsight's OpenTelemetry~\cite{opentelemetry} tracer.}
\vspace{-0.2in}
\label{tb:api}
\end{table}

\subsection{Client Library}

\autoref{tb:api} outlines Hindsight's client API.  Applications can interact with this API directly, or use Hindsight's OpenTelemetry tracer which acts as a wrapper.

\fakepara{Writing trace data.}
When a request begins executing in a thread, it must call \fun{begin}; subsequently it may call \fun{tracepoint} an arbitrary number of times; and finally when it completes executing in a thread, it must call \fun{end}.
This usage pattern is typical of distributed tracing frameworks.
\fun{tracepoint} accepts an arbitrary byte payload if called directly; conversely Hindsight's OpenTelemetry tracer serializes trace events as payload.
Hindsight internally maintains thread-local state including the current \traceid and a pointer to a buffer.
\fun{tracepoint} writes directly to the thread-local buffer without synchronization.
Synchronization is only required when acquiring or returning buffers; these operations touch shared-memory queues but are infrequent.
A buffer is acquired during \fun{begin}, returned during \fun{end}, and replaced when filled.

\fakepara{Communicating with agents.}
The client library acquires \bufferids by polling a shared-memory \emph{available queue}; if the queue is empty clients immediately return and instead write trace data to a special `null buffer' that is simply discarded.
When the client fills a buffer, it writes its \traceid and the \bufferid to a shared-memory \emph{complete queue}.
The agent continually drains the complete queue, and likewise continually returns fresh buffers to the available queue.
Shared memory queues are lock-free and support batch operations; using batch operations, agents are robust to queue contention from multiple client writer threads.

This paired channel design forms a natural separator between control and data with two desirable properties:
(1) queues only communicate metadata---a single integer \bufferid represents, by default, a 32\,kB buffer;
 (2) communication is infrequent, occurring only when buffers are filled or a thread switches over to execute a different request, thereby minimizing synchronization.
From the client library's perspective, it cheaply and blindly writes trace data into shared memory and forwards only the control metadata to agents; conversely agents are agnostic to buffer contents---they do not inspect data in the shared memory pool and use only the metadata communicated via the complete queue.

\fakepara{Depositing breadcrumbs.}
A breadcrumb is an address of a Hindsight agent.
When a request arrives at a node, it carries the breadcrumb of the previous node.
During trace context deserialization, the \traceid and breadcrumb is written to a shared memory breadcrumb queue.  Agents poll this queue and index breadcrumbs alongside buffer metadata.  Agents do not forward or act upon breadcrumbs until a trace is explicitly collected with a \trigger.
When a request departs a node, it takes that node's breadcrumb.  Clients can additionally establish forward-breadcrumbs to a named destination node prior to communication.
By following breadcrumbs, we can reconstruct the full request graph starting from any node, including for requests with arbitrary concurrency and fan-out.

\begin{table}[t]%
\rowcolors{1}{blue!3}{blue!9}%
\noindent%
\relsize{-1}%
\setlength{\tabcolsep}{2pt}%
\resizebox{\columnwidth}{!}{%
\begin{tabular}{p{0.28\columnwidth}p{0.67\columnwidth}}
\texttt{\textls[-100]{PercentileTrigger($p$)}} & Clients call \texttt{\textls[-75]{addSample(traceID, measurement)}}. Trigger fires for measurements \greaterthan percentile $p$.  (\eg{} high latency or resource consumption) \\
\texttt{\textls[-75]{CategoryTrigger($f$)}} & Clients call \texttt{\textls[-75]{addSample(traceID, label)}}. Trigger on categorical data that is less frequent than threshold $f$ (\eg{} rare API calls or attributes) \\
\fun{ExceptionTrigger} & Trigger on an exception or error code \\
\fun{TriggerSet($T$,$N$)} & Tracks the most recent $N$ \traceids and includes as \fun{lateralTraceIds} when $T$ fires. \\
\end{tabular}
}
\vspace{-2mm}
\caption{\textbf{Hindsight autotrigger API} can automatically trigger traces based on certain conditions.}%
\vspace{-0.2in}
\label{tb:autotrigger_api}%
\end{table}

\fakepara{Triggering trace collection.}
\label{sec:triggers}
\label{sec:implementation:autotriggers}
Applications initiate trace collection by invoking \trigger, which writes the \traceid, \triggerid and zero or more \fun{lateralTraceIds} to a shared-memory trigger queue.  In addition, Hindsight will propagate the fired trigger with the request similar to the \sampled flag (\cf~\autoref{fig:tracing_overview}) so that later nodes immediately learn of the trigger.

A developer can implement custom outlier detection and invoke \trigger directly, or they can make use of Hindsight's \autotrigger library (\autoref{tb:autotrigger_api}), a separate collection of triggers that track simple conditions over time and automatically invoke \trigger when a condition is met.
\fun{TriggerSet} is noteworthy as a building block for lateral tracing; it includes $N$ most recent traces whenever $T$ fires.

\subsection{Agent}

\fakepara{Trace index.} 
The trace index is a map of metadata, keyed by \traceid.
The metadata for a \traceid includes a list of \bufferids and a list of breadcrumbs.  Agents also maintain metadata of the \triggers that have fired.
Agents continually update the trace index with recently-written buffers, by polling \traceids and \bufferids from the complete queue.
The agent will evict traces when the index exceeds a threshold of buffer pool capacity (default 80\%) by
removing the least-recently used untriggered \traceid and returning all of its \bufferids to the available queue.

\fakepara{Local triggers.}
Agents poll the local trigger queue and immediately forward triggers to the coordinator.
Agents include the breadcrumbs of the triggered \traceid, enabling the coordinator to begin recursively disseminating the trigger to other agents.
Meanwhile the agent schedules the trigger to be reported.
In the case of a spammy local trigger, if the trigger exceeds a per-\triggerid rate-limit, the agent will immediately discard the trigger instead of forwarding and scheduling it.

\fakepara{Remote triggers.}
Agents receive remote triggers fired by other agents via the coordinator.
To facilitate rapid trigger dissemination, the agent immediately responds to a remote trigger by providing any breadcrumbs it has for the \traceid and \funn{lateralTraceIds}.
Unlike local triggers, agents do not rate-limit remote triggers---they are all  scheduled for reporting.

\fakepara{Reporting traces.}
When a trigger is scheduled for reporting, its \traceid and lateral \traceids are can no longer be evicted by the regular buffer eviction cycle.
The trigger is inserted into a per-\triggerid reporting queue.
In the normal case when an agent is not backlogged, the reporting queue will be empty.
The agent asynchronously pulls triggers from the queues; reads buffers of the \traceid and \funn{lateralTraceIds} from the buffer pool; 
sends the buffer contents to the backend collectors; and finally returns the \bufferids to the available queue.
A trace remains triggered even after reporting its data, in case the request is still generating trace data locally.

\fakepara{Ignoring triggers during overload.}
If the network or backend collectors are overloaded, reporting queues in an agent can fill up.
During overload, the agent continues to report traces as described above for the normal case.
The agent implements weighted fair queueing over the reporting queues and supports global and per-\triggerid reporting rate limits.
From a reporting queue, the agent dequeues the \emph{highest-priority trigger}, by using consistent hashing of \traceid, and reports its data as described above for the normal case.

Simultaneously, past a configured threshold, the agent must begin abandoning triggers to free up buffers.
Abandoning a trigger entails removing it from its reporting queue and returning buffers to the available queue.
Agents coherently select the \emph{lowest-priority trigger} to abandon, by using the same consistent hashing of \traceid.
In the case of multiple reporting queues, agents will ensure that a well-behaved \triggerid is not impacted by a spammy \triggerid:
agents implement weighted max-min fair-sharing across reporting queues to choose a queue from which to drop triggers.

\fakepara{Trigger priority ensures coherence during overload.}
Reporting queues are priority queues that use consistent hashing of \traceid to determine priority.
Across all agents, a given \traceid will enjoy the same priority relative to other \traceids.
Thus if multiple agents experience overload, they will coherently bias towards reporting the same high-priority \traceids and abandoning the same low-priority \traceids.

\section{Evaluation}
\label{sec:evaluation}

\newcommand{\notracing}{\textsc{\textls[-50]{None}}\xspace}
\newcommand{\othead}{\textsc{\textls[-50]{Head}}\xspace}
\newcommand{\ottail}{\textsc{\textls[-50]{Tail}}\xspace}
\newcommand{\ottailul}{\textsc{\textls[-50]{Tail}$^{\text{\hspace{-0.5mm}\textls[-200]{Sync}}}$}\xspace}
\newcommand{\hs}{\textsc{\textls[-50]{HS}}\xspace}
\newcommand{\hsul}{\textsc{\textls[-50]{HS}$^{\text{\hspace{-0.5mm}\textls[-200]{++}}}$}\xspace}
\newcommand{\hindsightgrpc}{MicroBricks\xspace}
\newcommand{\alibaba}{Ali\xspace}

We now evaluate how effectively Hindsight overcomes the fundamental problem of head-based tracing methods in examples (\ucA)--(\ucC) and meets the goals of retroactive sampling to provide lightweight and effective request tracing.

\fakepara{Systems.}  We evaluate Hindsight on three distributed systems.  To validate our motivating use cases (\ucA--\ucC), we integrate Hindsight with the Hadoop Distributed File System (HDFS)~\cite{hdfs}(with a $\approx$300LOC JNI-based Java client library) 
and the DeathStar Social Network Microservices Benchmark (DSB)~\cite{gan2019open}.  To assess Hindsight at greater scale and load, we develop a flexible, configurable RPC benchmark called \textbf{\hindsightgrpc}.

\hindsightgrpc is a microservice benchmark written in $\approx$3KLOC C\verb|++| using gRPC's high-performance \texttt{async} library.  A \hindsightgrpc deployment comprises a topology of RPC services such that each client request will traverse multiple services.  A call to a service will execute for some amount of time, then concurrently call zero or more other RPC services with some probability.  Each service is independently configured with its own set of APIs, each with their own execution times, child dependencies, and child call probabilities.  We evaluate using several different topologies.  In particular, we use Alibaba's microservice trace dataset~\cite{luo2021characterizing} to derive realistic topologies by calculating per-service execution time distributions, service dependencies, child call probabilities, and client workloads.

\fakepara{Baselines.}  We configure OpenTelemetry~\cite{opentelemetry} with Jaeger~\cite{jaeger} under head-sampling (1\% unless indicated) and tail-sampling.

\fakepara{Instrumentation.}  We instrument \hindsightgrpc with OpenTelemetry to create spans and events for RPC calls and child calls.  We use DSB's existing OpenTracing instrumentation and add support for Hindsight.  We use Hadoop's existing X-Trace instrumentation~\cite{fonseca2007xtrace} and update X-Trace to write its trace data to Hindsight.

\fakepara{Summary.} Our experiments demonstrate the following:
\newcommand\sbullet[1][.5]{\mathbin{\vcenter{\hbox{\scalebox{#1}{$\bullet$}}}}}
\begin{itemize}[noitemsep,topsep=0pt,parsep=0pt,partopsep=0pt,label={$\sbullet[.75]$},leftmargin=*]
\item Hindsight effectively addresses the overhead vs.~edge-cases trade-off faced by existing tracing frameworks.
\item Hindsight captures relevant edge-case traces across real use-cases (\ucA--\ucC).
\item Hindsight is lightweight and not a bottleneck for client applications, unlike OpenTelemetry~\cite{opentelemetry} and Jaeger~\cite{jaeger}.  Hindsight's trace API imposes nanosecond overheads; Hindsight's impact on end-to-end application latency and throughput is \lessthan3.5\% when tracing 100\% of requests and generating \greaterthan200\,MB/s of trace data per node.
\item Hindsight's control/data split provides up to 55\,GB/s write throughput.

\end{itemize}

\subsection{Overheads vs. Edge-Cases}
\label{sec:eval:overheads_vs_edgecases}

\begin{figure}
\centering%
\includegraphics[width=\linewidth,trim=2pt 2px 0 8px, clip]{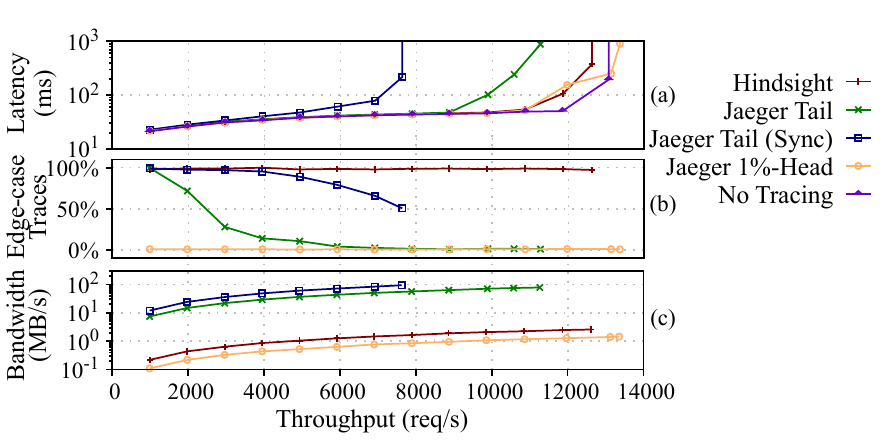}%
\vspace{-0.1in}
\caption{\textbf{Overhead vs.~edge-cases} on a 93-service Alibaba MicroBricks topology with 1\% edge-cases~(\autoref{sec:eval:overheads_vs_edgecases}).  For different tracing configurations we show: (a) application end-to-end latency-throughput curves; (b) the rate of coherent edge-trace cases captured; and (c) network bandwidth.}
\label{fig:overheads_vs_edgecases}
\label{fig:overheads_vs_edgecases:latency_throughput}
\label{fig:overheads_vs_edgecases:edgecases}
\label{fig:overheads_vs_edgecases:network}
\vspace{-0.2in}
\end{figure}

In this experiment, we evaluate Hindsight in a large-scale setting with a realistic microservice topology derived from Alibaba request traces~\cite{luo2021characterizing}.  We show that Hindsight overcomes the limitations of head-sampling and tail-sampling.  

We deploy \hindsightgrpc with a 93-service Alibaba topology in a 544-core private cluster (comprising 10$\times$Dell R920 48-core 1.5\,TB machines and 4$\times$Dell M620 16-core 256\,GB machines).  We deploy each service in a separate container.  We use separate machines to \textbf{(i)} generate workload and \textbf{(ii)} run the OpenTelemetry collector/Hindsight coordinator+collector.  

To directly control the number of edge-case traces, we randomly decide with low probability (1\%) to designate a request an edge-case when it completes (later experiments consider autotriggers).  We annotate the root span of edge-cases with an additional attribute so that tail-sampling can filter traces on this attribute.  Hindsight directly fires a trigger for edge-cases from within \hindsightgrpc.  We repeat the experiment multiple times, analyzing results under four tracing configurations:

\fakepara{Head Sampling (Jaeger 1\%-Head).}  \autoref{fig:overheads_vs_edgecases:latency_throughput}a shows the average request latency and throughput as we vary the offered load from 0 to 14,000 requests/sec (r/s).  Jaeger 1\%-Head has comparable peak throughput and latency as No Tracing, since it traces only 1\% of requests, thus amortizing the tracing overhead.  \autoref{fig:overheads_vs_edgecases:edgecases}b plots the percentage of coherent edge-case traces captured per second. Since head-sampling cannot discriminate, it only captures $\approx$1\% of all edge-case traces, peaking at 1.64 per second.  \autoref{fig:overheads_vs_edgecases:network}c shows the network bandwidth consumption between application nodes and the OpenTelemetry collector. With few requested being traced, Head-sampling only consumes a maximum of 1.4\,MB/s of network bandwidth.

\fakepara{Tail Sampling (Jaeger Tail).}  Tail-sampling imposes more burden on the traced application than head-sampling, attaining 14\% lower peak throughput (\autoref{fig:overheads_vs_edgecases:latency_throughput}a).  At low load (1,000\,r/s), tail-sampling successfully captures $\approx$100\% of edge-case traces, at 9.9 per second (\autoref{fig:overheads_vs_edgecases:edgecases}b).  However, a load of just 2,000\,r/s is sufficient for clients to encounter back-pressure from the network and the OpenTelemetry collector, and they begin incoherently dropping spans: at 2,000\,r/s only 71\% coherent edge-cases are captured; at 3,000\,r/s only 28\%; and so on.  Tail-sampling rapidly deteriorates and at peak load captures \emph{fewer} coherent edge-case traces than head-sampling (1.44\,edge-cases/s), because 98.8\% of captured traces are incoherent.  Tail-sampling consumes up to 78\,MB/s of network bandwidth (\autoref{fig:overheads_vs_edgecases:edgecases}c).

\fakepara{Tail Sampling (Jaeger Tail Sync).}  Jaeger clients asynchronously send spans to OpenTelemetry collectors, and as we just observed, drop spans when client-side queues fill up.  We repeat the experiment with a synchronous variant, whereby clients send spans to OpenTelemetry synchronously.  Back-pressure then manifests as additional critical-path request latency.  This approach inevitably increases request latency and reduces peak throughput by 42\% (\autoref{fig:overheads_vs_edgecases:latency_throughput}a).  However, we can observe the collector ultimately captures more edge-case traces, peaking at 47\,edge-cases per second at 6,000\,r/s (\autoref{fig:overheads_vs_edgecases:edgecases}b) and 72.2\,MB/s of network.  Beyond this, the OpenTelemetry collector is saturated and cannot process a higher rate of traces; it begins indiscriminately dropping incoming spans, reducing the fraction of coherent edge-case traces.

\fakepara{Hindsight.}  Hindsight achieves comparable peak throughput to No Tracing (\lessthan3.5\%), and minimal impact on request latency below peak load (\autoref{fig:overheads_vs_edgecases:edgecases}a).  Hindsight captures 99--100\% of edge-case traces at all throughputs (\autoref{fig:overheads_vs_edgecases:edgecases}b).  Hindsight consumes a maximum of 2.6\,MB/s of network bandwidth since only edge-case traces are being collected (\autoref{fig:overheads_vs_edgecases:edgecases}c).

\subsection{Scalability and Overload}
\label{sec:eval:scalability}

\newcommand{\ta}{\texttt{t}$_{\texttt{A}}$\xspace}
\newcommand{\tb}{\texttt{t}$_{\texttt{B}}$\xspace}
\newcommand{\tf}{\texttt{t}$_{\texttt{F}}$\xspace}
\newcommand{\taa}{\texttt{t}$_{\texttt{A}}$}
\newcommand{\tbb}{\texttt{t}$_{\texttt{B}}$}
\newcommand{\tff}{\texttt{t}$_{\texttt{F}}$}

\begin{figure}[t]
\centering
\includegraphics[width=\linewidth,trim=0 0 0 5px, clip]{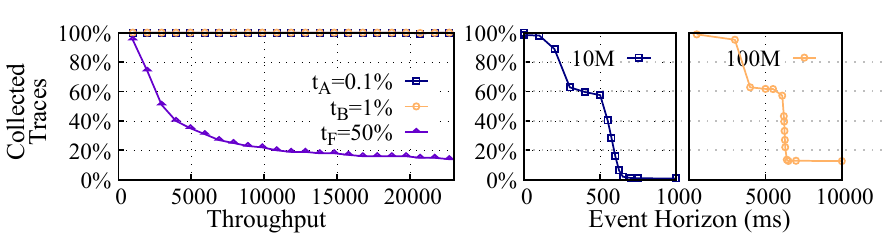}\\[-4mm]%
\begin{subfigure}[t]{0.5\linewidth}%
\caption{Coherent traces captured when overloaded with a spammy trigger \tf.}%
\label{fig:collected_traces}%
\end{subfigure}%
\hspace{0.02\linewidth}%
\begin{subfigure}[t]{0.48\linewidth}%
\caption{Event horizon for constrained bufferpools (10MB and 100MB).}%
\label{fig:event_horizon}%
\end{subfigure}\\[-3mm]%
\begin{subfigure}[t]{\linewidth}%
\includegraphics[width=\linewidth,trim=0 0 0 3px, clip]{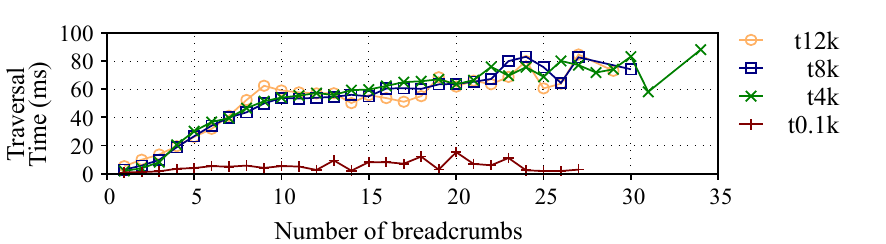}
\caption{Breadcrumb traversal time as trace size varies, for increasing trigger rates.}
\label{fig:breadcrumb_traversal}
\end{subfigure}%
\vspace{-0.1in}
\caption{\textbf{Scalability and Overload}  }
\label{fig:collected_traces_and_event_horizon}
\vspace{-0.2in}
\end{figure}

We now focus on two aspects of Hindsight's scalability: its breadcrumb traversal mechanism and its ability to rate-limit spammy triggers.  We deploy the 93-service Alibaba topology as described in~\autoref{sec:eval:overheads_vs_edgecases}.  To reach a higher request and trace throughput, we scale down the computation performed at each service and increase offered load up to 28,000\,r/s.  We install three triggers with probabilities \taa=0.1\%, \tbb=1\%, and \tff=50\%.  \tf represents a faulty trigger---it fires for 50\% of requests and thereby adds substantial load to Hindsight's breadcrumb traversal mechanism.  We rate-limit Hindsight's collector bandwidth to 1\,MB/s per agent to backlog the agents and inhibit Hindsight's ability to collect traces; thus \tf triggers far more traces than Hindsight can collect.

\fakepara{Coherent rate-limiting.} \autoref{fig:collected_traces} plots the percentage of coherent traces captured for \taa,\tbb{} and \tf as the offered load increases.  Throughout the experiment, Hindsight captures approximately 100\% of traces triggered by \ta and \tb, since they fire infrequently.  By contrast, \tf triggers far more traces than can be collected.  In absolute terms, Hindsight collects $\approx$2,000 coherent traces per second throughout the experiment, with \tf using capacity not used by \ta and \tb.  Thus, higher request rates results in more traces dropped for \tf in both relative and absolute terms.

\newcommand{\higha}{\texttt{\relsize{0}\textls[-100]{t12}k}\xspace}
\newcommand{\highb}{\texttt{\relsize{0}\textls[-75]{t8}k}\xspace}
\newcommand{\highc}{\texttt{\relsize{0}\textls[-75]{t4}k}\xspace}
\newcommand{\lowa}{\texttt{\relsize{0}\textls[-75]{t0\hspace{-0.3mm}.\hspace{-0.3mm}1}k}\xspace}

\begin{figure}[t]%
\centering%
\begin{subfigure}[b]{\linewidth}%
\centering%
\includegraphics[width=\linewidth,trim=0 0 0 5px, clip]{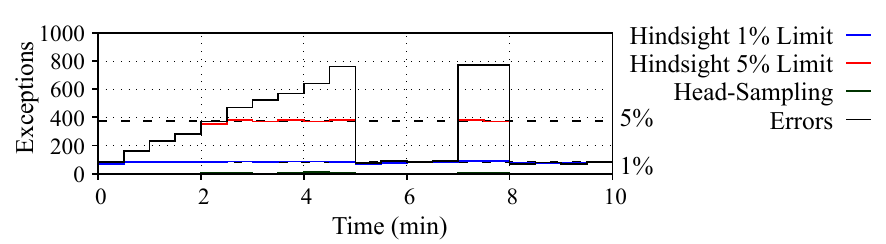}%
\vspace{-0.1in}
\caption{ \textbf{Error diagnosis (\ucA).} Exceptions captured by different sampling
strategies as the error rate varies.}
\label{fig:case_study_exception}
\end{subfigure}%
\\%
\begin{subfigure}[b]{\linewidth}%
\centering
\includegraphics[width=\linewidth,trim=0 0 0 3px, clip]{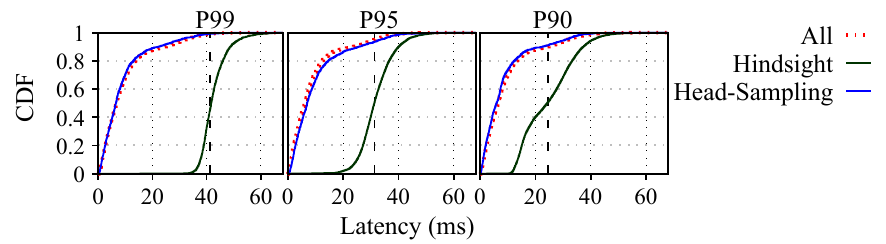}%
\vspace{-0.1in}
\caption{ \textbf{Tail-Latency (\ucB)}. Latency of requests captured through different sampling strategies with different tail-latency triggers (top to bottom).}
\label{fig:case_study_latency}
\end{subfigure}%
\\%
\begin{subfigure}[b]{\linewidth}%
\centering
\includegraphics[width=\linewidth,trim=0 0 0 5px, clip]{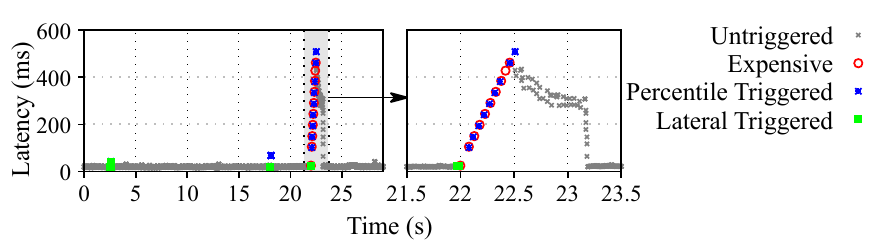}%
\vspace{-0.1in}
\caption{\textbf{Temporal Provenance (\ucC).} Lateral requests gathered (blue) after
triggering on slow requests (red) due to an overfull queue in HDFS.}
\label{fig:hadoop}
\end{subfigure}
\caption{\textbf{Hindsight applied on use cases \ucA--\ucC} (see~\autoref{sec:case_studies}).}
\vspace{-0.2in}
\end{figure}

\fakepara{Breadcrumb traversal.} \autoref{fig:breadcrumb_traversal} plots the average breadcrumb traversal time based on the trace size -- \ie the number of Hindsight agents that were recursively contacted.  We show results for four experiment iterations and label them based on their approximate trigger rates: \higha, \highb, and \highc correspond to the 24k, 16k and 8k\,r/s workloads ($\approx$12k, 8k, and 4k triggers/second respectively).  To compare to a non-overloaded setting we also include \lowa, a 12k\,r/s workload from~\autoref{sec:eval:overheads_vs_edgecases} ($\approx$0.1k triggers per second). Traversal time is elevated for \higha, \highb and \highc (up to 86\,ms) since spammy triggers substantially increase the load on Hindsight's coordinator.  Conversely, traversal time for \lowa is \lessthan13\,ms since triggers are relatively infrequent.  For each experiment, traversal time increases with trace size, but sub-linearly since breadcrumbs can be gathered concurrently from different branches in requests that have fan-out. However, even under the extremely overloaded circumstance, the longest traversal time is still manageable, which is less than 100\,ms and thus far smaller than the event horizon as described in the following section. 

\fakepara{Event horizon.}  We lastly measure Hindsight's event horizon. Here, we introduce a delay when an agent receives a local trigger.  We vary the delay added to triggers and measure how many coherent traces are ultimately collected.  At a certain point, triggers will have too much delay and trace data will have been evicted before the trigger even fires.  \autoref{fig:event_horizon} plots the percentage of coherent traces captured for \tb{} as we vary the trigger delay.  We repeat this experiment with small buffer pools (100\,MB and 10\,MB per agent) to exacerbate the event horizon effect.  Even a 10\,MB buffer pool can capture nearly 100\% coherent traces in the absence of added delays, but a 500\,ms delay drops coherence to 58\% and at 600\,ms, coherence is \lessthan 20\%.  A larger buffer pool improves the tolerance to delays: with a 100\,MB buffer pool, coherence surpasses 90\% with up to 3s delay, but drops to \lessthan 20\% by 6.4\,s. In practice, we believe our default 1\,GB pool is a reasonable choice, bringing an event horizon around 1 minute.

\subsection{Case Studies}
\label{sec:eval:casestudies}
\label{sec:hadoop}
We now turn our attention to the case studies introduced in~\autoref{sec:case_studies}, and demonstrate how Hindsight's local triggers are able to support these use cases.

\hyphenation{auto-trigger}
\fakepara{Error diagnosis (\ucA).}
We deploy DSB Social Network, a microservice system with 12 microservices and 17 backends~\cite{gan2019open}, on 13 CloudLab \texttt{c6320} nodes~\cite{duplyakin2019design}. We add an \fun{ExceptionTrigger} from Hindsight's autotrigger library to the ComposePostService, and run DSB's default workload with 300 r/s\footnote{We measure a maximum attainable DSB throughput of $\approx$350 r/s.}. We randomly inject exceptions in the ComposePostService module, with exception rates ranging from $1\%$ to $10\%$.  We repeat the experiment twice and rate-limit Hindsight's collector to approximately 1\% and 5\% of the total trace data generated by the experiment.
\autoref{fig:case_study_exception} plots the exception rate, and the number of coherent exceptional traces captured, for each 30\,s time window.  When there are few exceptions, Hindsight captures all traces; when the exception rate exceeds collector bandwidth, Hindsight coherently captures as many traces as possible within this limit.

\fakepara{Tail-latency (\ucB).}
We add a \fun{PercentileTrigger} from Hindsight's autotrigger library to the ComposePostService module in the same setting as above, invoking \fun{addSample} at the end of each ComposePost RPC call and providing the measured RPC duration.  We set $p$ to 99, 95, and 90, as different thresholds for tail latency. We inject $10\%$ requests at random with $20$--$30$\,ms latency. 
 \autoref{fig:case_study_latency} plots the latency distribution of requests captured by different strategies; the vertical dotted lines mark the tail-latency percentile threshold.  Hindsight is able to specifically target traces with high-percentile latency.  By contrast, head-sampling is random and thus its captured latency distribution resembles that of all requests -- useful for aggregate analysis but not for edge-case troubleshooting.  We note that Hindsight does not sacrifice this aggregate analysis use-case; it supports both simultaneously (\cf~\autoref{sec:eval:overheads_vs_edgecases}).

\fakepara{Temporal provenance (\ucC).}
We add a \fun{QueueTrigger} from Hindsight's \autotrigger library to the HDFS NameNode queue --- the \fun{QueueTrigger} combines a \fun{TriggerSet} with a \fun{PercentileTrigger}, parameterized to capture $N=10$ most recently dequeued lateral requests when 99.99$^\text{th}$ percentile queueing latency is observed.  We deploy HDFS on 10 machines (8 DataNodes, 1 NameNode, and 1 client) and run a Hindsight agent on each machine.  
We run a closed-loop workload of random 8\,kB reads with 10 concurrent requests.

\autoref{fig:hadoop} (left) shows NameNode queue latency over time. We inject a burst of 10 expensive \fun{createfile} requests 21 second into the trace that briefly saturate the queue---\autoref{fig:hadoop} (right) zooms in on this time window.  The figure shows high-latency requests (\textcolor{red}{\textbullet}), requests that fire the \autotrigger (\textcolor{blue}{\textbf{X}}), and the additional lateral requests that were triggered to Hindsight (\textbf{X}).  The first expensive request occurred at 22 seconds, followed by a pause while it was executed. Upon dequeuing the subsequent \fun{read8k} request, \fun{QueueTrigger} fired due to high queue latency, and Hindsight retroactively sampled the 10 prior traces leading up to the trigger. The sample included the culprit expensive request.  Overall, all 10 expensive requests were sampled, 8 unrelated requests prior to the first expensive request, and 9 additional \fun{read8k} requests.  Moreover, several intermittent latency spikes occurred unrelated to the experiment (\autoref{fig:hadoop}, left), which Hindsight also captured; upon investigation, these were due to garbage collection.

Unlike \ucA and \ucB, temporal provenance is unsupported in existing tail-samplers.  Moreover, temporal provenance is fundamentally difficult to support with tail-sampling due to scalability issues.  Temporal provenance requires knowledge of lateral traces (\eg the 10 previous traces); by implication those traces must all route to the same collector instance.  However in practice, tail-sampling necessarily uses \traceid for routing decisions -- thus related traces may arrive at different, oblivious collectors.

\begin{figure}%
\centering%
\includegraphics[width=\linewidth, trim=8px 6px 12px 7px, clip]{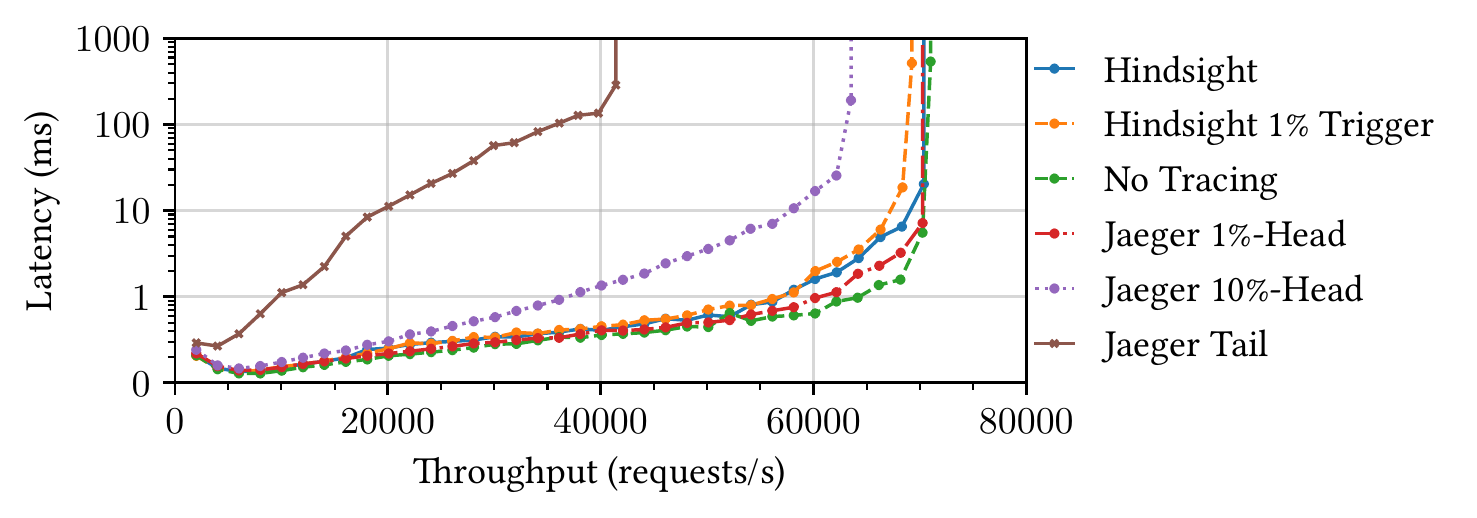}%
\vspace{-0.1in}
\caption{\textbf{End-to-end latency and throughput} for a 2-service \hindsightgrpc topology configured with various tracers, showing minimal application impact for Hindsight despite tracing 100\% of requests.}%
\label{fig:application_latency_throughput_microbenchmark}%
\vspace{-0.2in}
\end{figure}

\subsection{Hindsight Performance}
\label{sec:eval:tracing_performance}

\subsubsection{End-to-end Application Overheads}
\label{sec:eval:application_overhead}

Hindsight generates trace data for all requests; thus low overheads are a key goal of Hindsight's design.  In this experiment, we measure the impact of Hindsight on end-to-end application latency and throughput.  We deploy a two-service \hindsightgrpc topology with a 100\% call probability from the first service to the second.  To highlight tracing overheads, neither service performs additional compute.  We vary the offered load and measure end-to-end request latency and throughput.

\autoref{fig:application_latency_throughput_microbenchmark} plots latency-throughput curves under several different tracing configurations.  The lowest latency and highest throughput is achieved with No Tracing, peaking at an average 71.0\,k requests/s.  Similar throughput is achieved by Jaeger when configured with 1\% Head-sampling, at 70.2\,k r/s.  Hindsight peaks at 70.4\,k r/s -- a decrease of only 0.9\% compared to no tracing.  Hindsight generates on average 330\,MB/s of trace data at peak request throughput, with an event horizon of 5.2\,s, and consumes a combined 0.3 CPU cores across agents, coordinator, and collector.  By comparison, Jaeger configured with Tail-sampling peaks at only 41.4\,k r/s, an overhead 41.7\%; moreover, the workload over-saturates the OpenTelemetry collector, resulting in 94\% of trace data being dropped while consuming 4.5 CPU cores.

\subsubsection{Client API and Autotrigger Microbenchmarks}
\label{client_perf}
\label{sec:eval:client_perf}

\begin{table}%
\relsize{-1.5}%
\setlength{\tabcolsep}{1.5pt}%
\rowcolors{1}{}{}%
\begin{tabular}{@{}lrrr@{}r@{}lrrr@{}}
\cline{1-4} \cline{6-9}
\textbf{API Call} & \hspace{-5mm}\textls[-0]{\textbf{T=1}} & \textls[-0]{\textbf{T=4}} &\textls[-0]{\textbf{T=8}} && 
\textbf{API Call} & \textls[-0]{\textbf{T=1}} & \textls[-0]{\textbf{T=4}} &\textls[-0]{\textbf{T=8}} \\
\cline{1-4} \cline{6-9}
\hsbegin & 72.7 & 194.8 & 237.9 & \hspace{2mm} & \fun{\textls[-80]{tracepoint}} & 7.9 & 8.4 & 8.6\\
\hsend & 70.7 & 205.8 & 216.6 && \\
\fun{\textls[-70]{Category(.01)}} & 45.8 & 44.9 & 46.7 && \fun{\textls[-80]{tracepoint 8\,B}} & 3.9 & 4.0 & 4.8 \\
\fun{\textls[-100]{Percentile(99)}} & 275.3 & 293.5 & 306.9 && \fun{\textls[-80]{tracepoint 128\,B}} & 11.5 & 13.5 & 13.0 \\
\fun{\textls[-100]{Percentile(99.9)}} & 407.1 & 441.9 & 512.2 && \fun{\textls[-80]{tracepoint 512\,B}} & 37.7 & 43.1 & 40.9\\
\fun{\textls[-100]{Percentile(99.99)}} & 629.4 & 875.8 & 1134.0 && \fun{\textls[-80]{tracepoint 2\,kB}} & \hspace{-0.5mm}160.2 & 192.9 & 174.7\\
\fun{\textls[-70]{TriggerSet(10)}} & 6.57 & 44.1 & 52.2 \\
\cline{1-4} \cline{6-9}
\end{tabular}%
\vspace{-2mm}
\caption{Latency measurements (nanoseconds) for Hindsight client API and autotriggers for a microbenchmark application configured with \textbf{1}, \textbf{4}, and \textbf{8 Threads} (\autoref{sec:eval:client_perf}).  Default \tracepoint writes a 32\,kB trace event; we also measure 8--2048\,B \tracepoint payloads.}
\label{table:client_api_latency}
\vspace{-0.2in}
\end{table}

We run a benchmark application that generates traces and measures the overhead of calls to Hindsight's client API and autotrigger library.  The benchmark writes traces by calling \hsbegin to start the trace, writing a total of 16\,kB per trace by repeatedly calling \tracepoint, then calling \hsend to finish the trace.  Each \tracepoint call writes a 32-byte event struct (3 metadata fields and a timestamp) using Hindsight's OpenTelemetry library.  After each trace the benchmark invokes five different autotriggers.  The benchmark runs a configurable number of threads to generate traces; each thread runs a continuous loop generating traces, and each thread is independent and writes a different trace.
We configure Hindsight to use 32\,kB buffers and a 1\,GB buffer pool, and run a Hindsight agent.  We run 1 minute per experiment ($\approx$10--50 million traces).

\autoref{table:client_api_latency} shows API latency for 1, 4, and 8 threads.
Overall Hindsight achieves nanosecond-scale API latency, and by design the expensive API calls (\hsbegin, \hsend, and autotriggers) are limited to once per trace.  \hsbegin and \hsend vary from 70-230\,ns, proportional to the number of threads due to contending on shared-memory queues to acquire and return buffers.  By contrast, \tracepoint call latency is mostly independent of the number of threads, between 7.9--8.5\,ns (reduced to $\approx$4\,ns when omitting timestamps).  We also measure \tracepoint latency for larger payloads up to 2\,kB; latency increases only up to 175\,ns per tracepoint since \tracepoint is primarily a memory copy into the thread-local buffer established by \hsbegin.

Autotrigger overheads vary.  CategoryTrigger is relatively cheap (\lessthan47\,ns) and TriggerSet adds relatively little overhead to the wrapped trigger (6--53\,ns).  By contrast, PercentileTrigger overheads grow proportional to the percentile: up to $307$, $512$, and $1,134$\,ns respectively for tracking 99$^\text{th}$, 99.9$^\text{th}$, and 99.99$^\text{th}$ percentile latency.  This occurs due to larger internal data structures for tracking order statistics.

\subsubsection{Control-Data Trade-offs}
\label{sec:eval:control_data_tradeoffs}
Hindsight's design emphasizes a control-data split, to enable applications to write trace data at large volume while reducing the amount of indexing work agents must perform.  The main factor influencing this trade-off is Hindsight's buffer size.  With large buffers, agents index fewer buffers and thus perform less work; however it may exacerbate internal fragmentation when traces only partially fill buffers.  Conversely, small buffers are more space-efficient, but require more indexing work from agents.  We evaluate this trade-off by measuring client-side and agent-side throughputs, while varying Hindsight's internal buffer size from very small (128\,B) to very large buffers (128\,kB).

\begin{figure}[t]%
\centering%
\includegraphics[height=0.42\linewidth, trim=11px 0 80px 10px, clip]{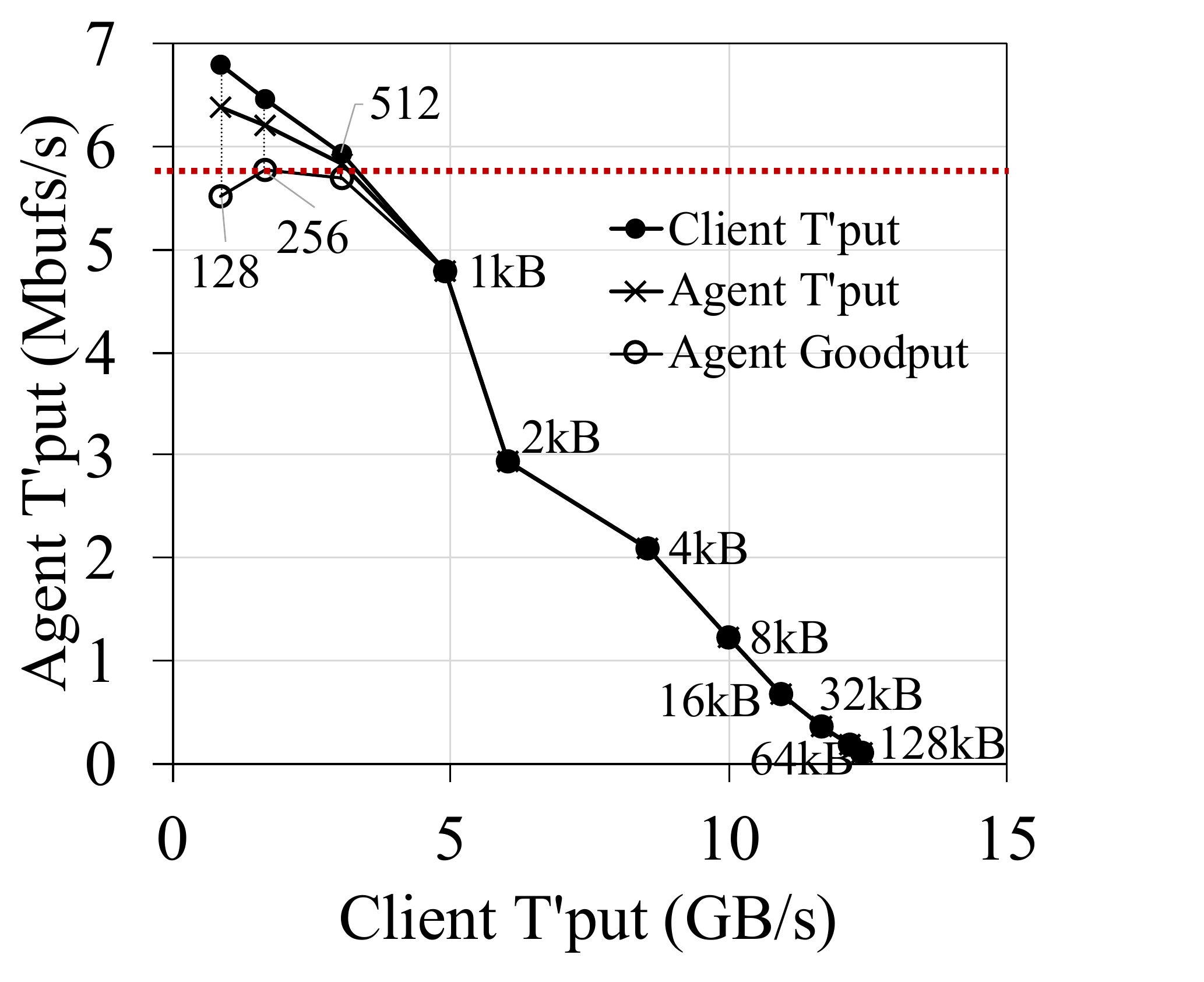}%
\includegraphics[height=0.42\linewidth, trim=20px 0 0 10px, clip]{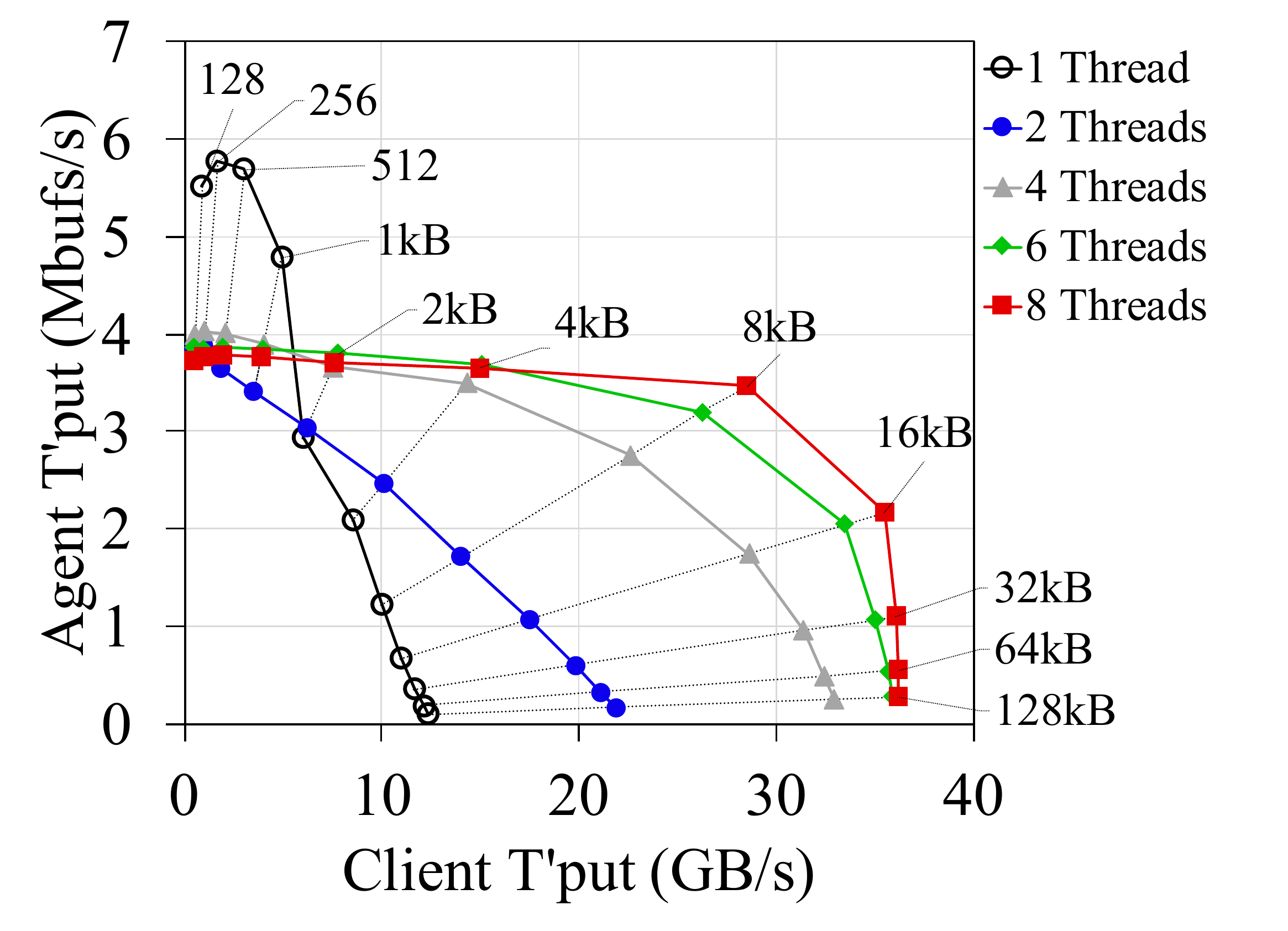}%
\vspace{-0.1in}
\caption{\textbf{Buffer size trade-off.}  Each data point is annotated with the Hindsight buffer size.  Small buffers require more indexing work from agents, while large buffers are less memory efficient by exacerbating internal fragmentation.}
\vspace{-0.2in}
\label{fig:throughput_scatter}
\end{figure}

We run the benchmark application with one thread, 100\,kB traces, and a payload of 1\,kB per \tracepoint call (Hindsight fragments payloads across multiple buffers when necessary).
\autoref{fig:throughput_scatter} (left) plots the client-side throughput of generating data ($x$-axis) and the agent-side throughput of indexing buffers ($y$-axis).  We annotate data points with the corresponding buffer size used.
Large buffer sizes (128\,kB) can support peak client data throughput (12.1\,GB/s) while requiring little of the agent.  Conversely, tiny buffer sizes (128\,B) stress the agent buffer throughput since we more frequently cycle buffers through the queues. \autoref{fig:throughput_scatter} (left) plots three lines and indicates two important phenomena.  The client throughput line plots the rate at which the client writes buffers, whereas the agent throughput line plots the rate at which the agent cycles buffers; the delta in-between are `null buffers', written by the client because the available queue is empty, \ie the agent cannot keep up.  Writing to null buffers means lost trace data; the third line, agent goodput, only counts buffers of coherent traces, \ie excluding all buffers for traces that lost data.  We observe that the goodput with 128\,B buffers is lower than with 256\,B buffers due to greater loss. In general, with {\relsize{-1}{$\geq$}}1\,kB buffers, the agent is able to consistently keep up without losing data.

\autoref{fig:throughput_scatter} repeats this experiment with varying numbers of threads, and plots client-side data throughput and agent-side buffer goodput.  
Buffer sizes of 16\,kB and higher are sufficient for reaching peak write throughput while remaining comfortably within agent throughput limits; by default, we select 32\,kB for Hindsight.

\section{Discussion}
\label{sec:discussion}

In this section we provide additional discussion that is peripheral to the core of Hindsight's design.

\subsection{Failures}

\fakepara{Application Failures.} If the application process crashes (e.g. SEGV/NPE-type crashes), then Hindsight will be able to preserve problematic traces, because Hindsight's agent will continue to run and the trace data will be preserved in memory in the shared buffer pool.  The agent will also be able to continue responding to breadcrumbs.  This is a secondary benefit of externalizing trace data on the critical path of requests, and Hindsight currently supports this.  By contrast existing distributed tracing frameworks buffer trace data in application memory and would lose unreported data upon an application crash.

\fakepara{Agent Failures.} If Hindsight's agent crashes (including or excluding the application process crashing), then the buffer pool will still exist in-memory on the machine and could be later retrieved to inspect the state just prior to the crash.  Hindsight does not currently implement such a recovery process.  In addition, if an agent crashes it will by default prevent Hindsight's coordinator from following breadcrumbs through this crashed agent.  This can be overcome with a straightforward extension of Hindsight's breadcrumb mechanism -- propagating breadcrumbs for the last N visited nodes instead of just one; this would both avoid (N-1)-hop failures and also speed up Hindsight's collection process.

\fakepara{Kernel and Hardware Failures.} In the case of kernel crashes or hardware failure, application-level traces are only useful if it was the application's behavior that triggered the crash.  In this case Hindsight's data would be lost.

\subsection{Cross-Layer Tracing} 

\fakepara{Cross-layer telemetry data.} Distributed tracing is primarily useful at troubleshooting application-level problems in distributed systems, because all data is generated from within the application and the tracing frameworks are thus limited to only data visible to the application.  Cross-layer and lower-layer issues are a challenge in their own right and an open research problem (e.g. recent work specifically focusing on which layer to attribute a problem too~\cite{gao2022buffer, ardelean2018performance}).  The primary challenge is figuring out how to integrate multiple different sources of telemetry data at different levels, of which distributed tracing is just one.  This is a focus of our ongoing work, since Hindsight is designed to handle high volumes of telemetry data, thus making it feasible to cheaply integrate other detailed cross-layer sources of data into Hindsight as a single point of ingestion.

\fakepara{Detecting symptoms from other layers.}  Hindsight can only persist traces for symptoms that can be observed at the application level.  In cases such as short-lived network congestion, Hindsight can't detect this at the network level.  However, if congestion bubbles up as an application-level SLO violation, then Hindsight would be able to detect it and fire a trigger.  Recent work has embedded application trace IDs in network packets~\cite{gao2022buffer}, and we are considering this in our ongoing work as a way to externalize triggers, e.g. network-level triggers can fire and persist application-level traces.

\subsection{Event Horizon}

\fakepara{Event Horizon Factors.}  Several factors influence Hindsight's event horizon: (i) the bufferpool size of each agent; (ii) the rate new trace data is generated; (iii) the time between a request completing and a trigger firing.  Inevitably, if there is too much trace data, or if triggers are too slow, Hindsight may be unable to keep the trace before its data is overwritten.  For some use cases this means Hindsight cannot use retroactive sampling.  However, head-sampling or tail-sampling would still be viable options, equivalent to existing distributed tracing frameworks.

\fakepara{Increasing the Event Horizon.}  The solution is either to increase the memory available to Hindsight or to scale down the percentage of traced requests using Hindsight's optional \emph{trace percentage}.  Trace percentage is a separate configuration knob (defaulting to 100\%) that controls the percentage of requests that generate trace data in the first place.  The starting premise for Hindsight is that 100\% tracing is acceptable, so we use 100\% as the default and describe Hindsight as such throughout the paper.  However this is not mandatory.  If an application has overhead constraints or limited memory for bufferpool, then it can scale-back the percentage of requests that are traced in the first place.
Hindsight enforces scale-back coherently across agents through consistent \traceid hashing, and reducing the trace percentage has a corresponding increase in event horizon, \eg 50\% trace percentage will half trace data throughput and double event horizon.

\fakepara{Mismatched and Dynamic Event Horizons.}  The global event horizon of an application is dictated by the shortest event horizon among the constituent processes, because the moment the first agent evicts the data of a trace, that whole trace becomes incoherent.  This is a fundamental property of Hindsight, and one that can primarily be addressed by configuring larger bufferpool memory on higher throughput nodes.  The bufferpool does not need fundamentally need to be fixed-size and we considered implementing a dynamically-sized bufferpool, \eg that can be configured with a target event horizon.  Ultimately we chose a fixed-size bufferpool to better bound memory overheads, a desirable property for telemetry systems~\cite{verma2015borg}.

\fakepara{Shared Buffer Pools.}  In our current design we deploy 1 Hindsight agent per traced application process.  If multiple containers share a machine, as in our experiments, then this results in multiple agents running on the same machine.  In principle there is no reason applications cannot share a single machine-wide bufferpool.  Doing this would allow processes to pool their bufferpool capacity and it would average-out any difference in event horizon between the processes.

\subsection{Event Horizon for Tail Sampling}

\fakepara{Tail Sampling Event Horizons.}  Hindsight's event horizon has an analogue in tail sampling, because trace collectors cannot immediately perform tail sampling the instance trace data arrives.  Instead a collector must wait for all of the slices of a trace to arrive from all of the machines the request visited.  Today this is done with a timeout (\eg 30s by default in OpenTelemetry~\cite{opentelemetry}), after which the trace objects are constructed and tail samplers can be evaluated.  If the application generates a high volume of trace data, then the trace collector can potentially run out of memory buffering data while waiting to do tail sampling.

\fakepara{Tail Sampling Expressivity.}  Today's tail samplers focus on filters and outliers applied to span attributes and metrics.  More fundamentally, a tail sampling decisions for one trace cannot influence the sampling decision of other traces.  By contrast, Hindsight's lateral traces enable a trigger to specify other, related traces, in addition to the one exhibiting a symptom.  Thus Hindsight can support use cases like temporal provenance (\ucC).  Tail sampling does not support such use cases, and it would be non-trivial to introduce this ability, due to the way trace data for different traces route to different collectors based on \traceid.

\section{Related Work}
\label{sec:related}

\fakepara{Distributed tracing.}
Numerous prior works identify end-to-end requests as a useful granularity for slicing telemetry data and troubleshooting distributed systems. Example use cases include detecting anomalous request structures~\cite{wu2019zeno, sigelman2010dapper, las2019sifter}, diagnosing changes in the steady-state~\cite{chen2004path, sambasivan2011diagnosing, ostrowski2011diagnosing}, modeling workloads~\cite{thereska2006stardust, mann2011modeling}, and identifying resource and queue contention~\cite{wu2019zeno, mace2015retro, gan2019seer}.  Distributed tracing systems have been presented in industry~\cite{sigelman2010dapper, kaldor2017canopy}, as open-source tools~\cite{jaeger, zipkin, opentracing, opentelemetry}, and in academia~\cite{fonseca2007xtrace, mace2015pivot}.  Edge-case troubleshooting stands in tension with overheads in distributed tracing, and head-sampling and tail-sampling offer alternative points in this space (\autoref{sec:motivation:distributed_tracing}).

\fakepara{Logging frameworks.}
Distributed tracing is the cousin of log ingestion frameworks that collect and store application-level log data~\cite{boulon2008chukwa, splunkLoggingPerformance}.  Log ingestion frameworks are agnostic to concepts like requests, do not record or group log data by requests, and cannot control head-sampling decisions coherently for requests -- instead applications generate simple sequential streams of log data all at the same level of logging detail.  Consequently, logs are typically far less detailed than distributed tracing and log ingestion frameworks handle a lower volume of data.  For example Chukwa reports on average 10kB/s per node~\cite{boulon2008chukwa}; Splunk limits to 330 kB/s per node~\cite{splunkLoggingPerformance}; Amazon CloudWatch limits to 5MB/s per log stream~\cite{cloudwatchLogsQuotas}.  Early distributed tracing works rejected the idea of building distributed tracing atop logging, citing coherence challenges from brittle data, enormous post-processing costs, and fundamental scalability bottlenecks~\cite{sigelman2010dapper, kaldor2017canopy, chow2014mystery}.  In practice, trace detail is typically far greater than even non-production debug-level logging~\cite{sigelman2010dapper}, and it is easy to see why: head-sampling gives operators leeway to instrument their applications at fine detail, because they can amortize the high cost of a single trace by scaling down the number of collected traces.  By comparison, log ingestion frameworks have no such opportunity.

\fakepara{Network provenance.}
Hindsight is similar in spirit to network packet provenance systems that chronicle the history of network state, enabling use cases such as tracking the origin or path traversed by a packet across the network. Earlier systems, like ExSPAN~\cite{zhou2010efficient} and SNP~\cite{zhou2011secure}, adopt this abstraction; more recent works like SyNDB~\cite{kannan2021debugging} and SPP~\cite{chen2017one} apply network provenance for packet-level root-cause analysis on Internet scale.
Packet provenance systems primarily trace only packet metadata, which is well-structured and can be summarized in-band; these systems tackle additional trust challenges outside of Hindsight's purview.
By contrast, handling metadata to reconstruct the path of a trace is but one concern for Hindsight; Hindsight is focused on handling arbitrary payloads (\ie trace data), and the resulting performance, coherence, and fairness challenges.  
Hindsight also draws inspiration from works focused on temporal provenance~\cite{zhou2012distributed} and packet reputation~\cite{chen16good} in distributed systems, although Hindsight's tracing abstractions operate entirely at the application level.

\section{Conclusion}

Hindsight circumvents the false dilemma %
between overhead and usefulness for diagnosing symptomatic edge cases by providing developers detailed 
traces from the recent past when they encounter symptoms of failures.
We believe the retroactive sampling abstraction, and our Hindsight implementation of it, can shift the conversation around tracing away from mechanism (how to collect traces) to a question of policy (what traces should be collected), and allow distributed tracing
systems to support edge-cases analysis: a key use case for which they were originally conceived.

\newpage

\bibliographystyle{plain}
\bibliography{ref}

\begin{thebibliography}{10}

\bibitem{businesslosingindowntime}
{Businesses Losing \$700 Billion a Year to IT Downtime, Says IHS}.
\newblock Retrieved April 2022 from
  \url{https://www.businesswire.com/news/home/20160125005188/en/Businesses-Losing-700-Billion-a-Year-to-IT-Downtime-Says-IHS}.

\bibitem{awsoutage}
{Recent AWS outage and how you could have avoided downtime}.
\newblock Retrieved April 2022 from
  \url{https://medium.com/@datapath_io/recent-aws-outage-and-how-you-could-have-avoided-downtime-7d9d9443d776}.

\bibitem{hdfs3751}
{HDFS-3751: DN should log warnings for lengthy disk IOs}.
\newblock Retrieved April 2022 from
  \url{https://issues.apache.org/jira/browse/HDFS-3751}, 2014.

\bibitem{hbase8228}
{HBASE-8228: Investigate time taken to snapshot memstore}.
\newblock Retrieved April 2022 from
  \url{https://issues.apache.org/jira/browse/HDFS-8228}, 2015.

\bibitem{hbase8744}
{HBASE-8744: Enable HBase to log the entire latency profile for HDFS packets
  resulting in slow writes.}
\newblock Retrieved April 2022 from
  \url{https://issues.apache.org/jira/browse/HDFS-8744}, 2016.

\bibitem{hdfs11461}
{HDFS-11461: DataNode Disk Outlier Detection}.
\newblock Retrieved April 2022 from
  \url{https://issues.apache.org/jira/browse/HDFS-11461}, 2017.

\bibitem{jaeger425}
{Jaeger Issue 425: Discuss post-trace (tail-based) sampling}.
\newblock Retrieved April 2022 from
  \url{https://github.com/jaegertracing/jaeger/issues/425}, 2017.

\bibitem{hdfs6110}
{HDFS-6110: adding more slow action log in critical write path}.
\newblock Retrieved April 2022 from
  \url{https://issues.apache.org/jira/browse/HDFS-6110}, 2018.

\bibitem{jaeger1861}
{Jaeger Issue 1861: Delayed Sampling}.
\newblock Retrieved April 2022 from
  \url{https://github.com/jaegertracing/jaeger/issues/1861}, 2019.

\bibitem{grafanaHorizontalScalability}
{Annanay Agarwal}.
\newblock {How Grafana Labs enables horizontally scalable tail sampling in the
  OpenTelemetry Collector}.
\newblock Retrieved April 2022 from
  \url{https://grafana.com/blog/2020/06/18/how-grafana-labs-enables-horizontally-scalable-tail-sampling-in-the-opentelemetry-collector/},
  2020.

\bibitem{ardelean2018performance}
Dan Ardelean, Amer Diwan, and Chandra Erdman.
\newblock Performance analysis of cloud applications.
\newblock In {\em 15th USENIX Symposium on Networked Systems Design and
  Implementation (NSDI'18)}, pages 405--417, 2018.

\bibitem{zipkinSecondarySampling}
Narayanan Arunachalam.
\newblock {Zipkin Secondary Sampling}.
\newblock Retrieved April 2022 from
  \url{https://github.com/openzipkin-contrib/zipkin-secondary-sampling}, 2019.

\bibitem{cloudwatchLogsQuotas}
{AWS}.
\newblock {AWS CloudWatch Logs quotas}.
\newblock Retrieved April 2022 from
  \url{https://docs.aws.amazon.com/AmazonCloudWatch/latest/logs/cloudwatch_limits_cwl.html},
  2022.

\bibitem{boulon2008chukwa}
Jerome Boulon, Andy Konwinski, Runping Qi, Ariel Rabkin, Eric Yang, and Mac
  Yang.
\newblock Chukwa, a large-scale monitoring system.
\newblock In {\em Proceedings of CCA}, volume~8, pages 1--5, 2008.

\bibitem{chen2017one}
Ang Chen, Andreas Haeberlen, Wenchao Zhou, and Boon~Thau Loo.
\newblock One primitive to diagnose them all: Architectural support for
  internet diagnostics.
\newblock In {\em Proceedings of the Twelfth European Conference on Computer
  Systems}, pages 374--388, 2017.

\bibitem{chen16good}
Ang Chen, Yang Wu, Andreas Haeberlen, Wenchao Zhou, and Boon~Thau Loo.
\newblock The good, the bad, and the differences: Better network diagnostics
  with differential provenance.
\newblock In {\em Proceedings of the 2016 ACM SIGCOMM Conference}, SIGCOMM '16,
  page 115–128, New York, NY, USA, 2016. Association for Computing Machinery.

\bibitem{chen2004path}
Mike~Y Chen, Anthony Accardi, Emre Kiciman, Jim Lloyd, Dave Patterson, Armando
  Fox, and Eric Brewer.
\newblock Path-based failure and evolution management.
\newblock In {\em 1st USENIX Symposium on Networked Systems Design \&
  Implementation (NSDI'04)}, pages 23--23, 2004.

\bibitem{chow2014mystery}
Michael Chow, David Meisner, Jason Flinn, Daniel Peek, and Thomas~F Wenisch.
\newblock {The Mystery Machine: End-to-end Performance Analysis of Large-scale
  Internet Services}.
\newblock In {\em 11th USENIX Symposium on Operating Systems Design and
  Implementation (OSDI'14)}, 2014.

\bibitem{dean2013tail}
Jeffrey Dean and Luiz~Andr{\'e} Barroso.
\newblock {The Tail at Scale}.
\newblock {\em Communications of the ACM}, 56(2):74--80, 2013.

\bibitem{40801}
Jeffrey Dean and Luiz~André Barroso.
\newblock The tail at scale.
\newblock {\em Communications of the ACM}, 56:74--80, 2013.

\bibitem{duplyakin2019design}
Dmitry Duplyakin, Robert Ricci, Aleksander Maricq, Gary Wong, Jonathon Duerig,
  Eric Eide, Leigh Stoller, Mike Hibler, David Johnson, Kirk Webb, et~al.
\newblock The design and operation of {CloudLab}.
\newblock In {\em 2019 USENIX Annual Technical Conference (USENIX ATC'19)},
  pages 1--14, 2019.

\bibitem{elasticco}
{elastic}.
\newblock {Transaction Sampling}.
\newblock Retrieved April 2022 from
  \url{https://www.elastic.co/guide/en/apm/guide/current/sampling.html#sampling}.

\bibitem{erlingsson2012fay}
{\'U}lfar Erlingsson, Marcus Peinado, Simon Peter, Mihai Budiu, and Gloria
  Mainar-Ruiz.
\newblock Fay: Extensible distributed tracing from kernels to clusters.
\newblock {\em ACM Transactions on Computer Systems (TOCS)}, 30(4):1--35, 2012.

\bibitem{fonseca2007xtrace}
Rodrigo Fonseca, George Porter, Randy~H Katz, and Scott Shenker.
\newblock X-trace: A pervasive network tracing framework.
\newblock In {\em 4th USENIX Symposium on Networked Systems Design \&
  Implementation (NSDI'07)}, 2007.

\bibitem{gan2019open}
Yu~Gan, Yanqi Zhang, Dailun Cheng, Ankitha Shetty, Priyal Rathi, Nayan Katarki,
  Ariana Bruno, Justin Hu, Brian Ritchken, Brendon Jackson, et~al.
\newblock An open-source benchmark suite for microservices and their
  hardware-software implications for cloud \& edge systems.
\newblock In {\em Proceedings of the 24th International Conference on
  Architectural Support for Programming Languages and Operating Systems
  (ASPLOS'19)}, pages 3--18, 2019.

\bibitem{gan2019seer}
Yu~Gan, Yanqi Zhang, Kelvin Hu, Dailun Cheng, Yuan He, Meghna Pancholi, and
  Christina Delimitrou.
\newblock Seer: Leveraging big data to navigate the complexity of performance
  debugging in cloud microservices.
\newblock In {\em Proceedings of the 24th International Conference on
  Architectural Support for Programming Languages and Operating Systems
  (ASPLOS'19)}, pages 19--33, 2019.

\bibitem{gao2022buffer}
Kaihui Gao, Chen Sun, Shuai Wang, Dan Li, Yu~Zhou, Hongqiang~Harry Liu, Lingjun
  Zhu, and Ming Zhang.
\newblock {Buffer-based End-to-end Request Event Monitoring in the Cloud}.
\newblock In {\em 19th USENIX Symposium on Networked Systems Design and
  Implementation (NSDI 22)}, 2022.

\bibitem{honeycombWhitepaper}
honeycomb.io.
\newblock {Getting At The Good Stuff: How To Sample Traces in Honeycomb}.
\newblock Technical report, honeycomb.io, 2019.

\bibitem{huang2018capturing}
Peng Huang, Chuanxiong Guo, Jacob~R Lorch, Lidong Zhou, and Yingnong Dang.
\newblock Capturing and enhancing in situ system observability for failure
  detection.
\newblock In {\em 13th USENIX Symposium on Operating Systems Design and
  Implementation (OSDI 18)}, 2018.

\bibitem{intelpt}
{Intel Corporation}.
\newblock {\em {Intel 64 and IA-32 architectures software developer’s
  manual}}, volume 3 (3A, 3B, 3C \& 3D): System Programming Guide.
\newblock Intel, 2016.

\bibitem{honeycombSampling}
{Irving Popovetsky}.
\newblock {Getting At The Good Stuff: How To Sample Traces in Honeycomb}.
\newblock Retrieved April 2022 from
  \url{https://www.honeycomb.io/blog/getting-at-the-good-stuff-how-to-sample-traces-in-honeycomb/},
  2020.

\bibitem{kamonIncoherentEdgecases}
{Ivan Topolnjak}.
\newblock {Kamon: How to Keep Traces for Slow and Failed Requests.}
\newblock Retrieved April 2022 from
  \url{https://kamon.io/blog/how-to-keep-traces-for-slow-and-failed-requests/},
  2021.

\bibitem{jaeger}
{Jaeger: Open Source, End-to-End Distributed Tracing}.
\newblock Retrieved April 2022 from \url{https://www.jaegertracing.io/}.

\bibitem{newstackTracing}
{Jeremy Castile}.
\newblock {What You Need to Know About Distributed Tracing and Sampling}.
\newblock Retrieved April 2022 from
  \url{https://thenewstack.io/what-you-need-to-know-about-distributed-tracing-and-sampling/},
  2020.

\bibitem{srebook}
Chris Jones, John Wilkes, Niall Murphy, and Cody Smith.
\newblock {\em {Site Reliability Engineering: How Google Runs Production
  Systems}}.
\newblock O'Reilly Media, 2016.
\newblock
  \url{https://landing.google.com/sre/sre-book/chapters/service-level-objectives/}.

\bibitem{kaldor2017canopy}
Jonathan Kaldor, Jonathan Mace, Micha{\l} Bejda, Edison Gao, Wiktor Kuropatwa,
  Joe O'Neill, Kian~Win Ong, Bill Schaller, Pingjia Shan, Brendan Viscomi,
  et~al.
\newblock Canopy: An end-to-end performance tracing and analysis system.
\newblock In {\em Proceedings of the 26th ACM Symposium on Operating Systems
  Principles (SOSP'17)}, pages 34--50, 2017.

\bibitem{kannan2021debugging}
Pravein~Govindan Kannan, Nishant Budhdev, Raj Joshi, and Mun~Choon Chan.
\newblock Debugging transient faults in data centers using synchronized
  network-wide packet histories.
\newblock In {\em 18th {USENIX} Symposium on Networked Systems Design and
  Implementation ({NSDI} 21)}, pages 253--268. {USENIX} Association, April
  2021.

\bibitem{lascasas2018weighted}
Pedro Las-Casas, Jonathan Mace, Dorgival Guedes, and Rodrigo Fonseca.
\newblock {Weighted Sampling of Execution Traces: Capturing More Needles and
  Less Hay}.
\newblock In {\em 9th ACM Symposium on Cloud Computing (SOCC '18)}, 2018.

\bibitem{las2019sifter}
Pedro Las-Casas, Giorgi Papakerashvili, Vaastav Anand, and Jonathan Mace.
\newblock Sifter: Scalable sampling for distributed traces, without feature
  engineering.
\newblock In {\em Proceedings of the ACM Symposium on Cloud Computing
  (SOCC'19)}, pages 312--324, 2019.

\bibitem{li2014tales}
Jialin Li, Naveen~Kr Sharma, Dan~RK Ports, and Steven~D Gribble.
\newblock {Tales of the Tail: Hardware, OS, and Application-Level Sources of
  Tail Latency}.
\newblock In {\em Proceedings of the 5th ACM Symposium on Cloud Computing
  (SoCC)}, 2014.

\bibitem{lightstepMicrosatellites}
{Lightstep}.
\newblock {Learn about Microsatellites: How many Microsatellites do I need?}
\newblock Retrieved April 2022 from
  \url{https://docs.lightstep.com/docs/learn-about-micro-satellites}.

\bibitem{ot407}
{Lightstep}.
\newblock {OpenTelemetry-Collector Issue \#4758: Tail-Based Sampling
  Scalability Issues}.
\newblock Retrieved April 2022 from
  \url{https://github.com/open-telemetry/opentelemetry-collector-contrib/issues/4758},
  2020.

\bibitem{luo2018troubleshooting}
Liang Luo, Suman Nath, Lenin~Ravindranath Sivalingam, Madan Musuvathi, and Luis
  Ceze.
\newblock Troubleshooting transiently-recurring errors in production systems
  with blame-proportional logging.
\newblock In {\em 2018 USENIX Annual Technical Conference (USENIX ATC'18)},
  pages 321--334, 2018.

\bibitem{luo2021characterizing}
Shutian Luo, Huanle Xu, Chengzhi Lu, Kejiang Ye, Guoyao Xu, Liping Zhang,
  Yu~Ding, Jian He, and Chengzhong Xu.
\newblock Characterizing microservice dependency and performance: Alibaba trace
  analysis.
\newblock In {\em Proceedings of the ACM Symposium on Cloud Computing}, pages
  412--426, 2021.

\bibitem{luo2022hubble}
Yu~Luo, Kirk Rodrigues, Lijin Jiang, Bing Xia, David Lion, and Ding Yuan.
\newblock {Hubble: Performance Debugging with In-Production, Just-In-Time
  Method Tracing on Android}.
\newblock In {\em 15th USENIX Symposium on Operating Systems Design and
  Implementation (OSDI 22)}, 2022.

\bibitem{mace2015retro}
Jonathan Mace, Peter Bodik, Rodrigo Fonseca, and Madanlal Musuvathi.
\newblock {Retro: Targeted Resource Management in Multi-Tenant Distributed
  Systems}.
\newblock In {\em Proceedings of the 12th USENIX Symposium on Networked Systems
  Design and Implementation (NSDI '15)}, 2015.

\bibitem{mace2015pivot}
Jonathan Mace, Ryan Roelke, and Rodrigo Fonseca.
\newblock {Pivot Tracing: Dynamic Causal Monitoring for Distributed Systems}.
\newblock In {\em 25th {ACM} Symposium on Operating Systems Principles
  (SOSP'15)}, 2015.

\bibitem{mann2011modeling}
Gideon Mann, Mark Sandler, Darja Krushevskaja, Sudipto Guha, and Eyal Even-Dar.
\newblock Modeling the parallel execution of black-box services.
\newblock In {\em Proceedings of USENIX Workshop on Hot Topics in Cloud
  Computing (HotCloud'11)}, 2011.

\bibitem{misra2019managing}
Pulkit~A Misra, Mar{\'\i}a~F Borge, {\'I}{\~n}igo Goiri, Alvin~R Lebeck, Willy
  Zwaenepoel, and Ricardo Bianchini.
\newblock Managing tail latency in datacenter-scale file systems under
  production constraints.
\newblock In {\em Proceedings of the Fourteenth EuroSys Conference 2019}, pages
  1--15, 2019.

\bibitem{newrelicTailsamplingUsecases}
{New Relic}.
\newblock {Technical distributed tracing details: Tail-based sampling
  algorithms}.
\newblock Retrieved April 2022 from
  \url{https://docs.newrelic.com/docs/distributed-tracing/concepts/how-new-relic-distributed-tracing-works/#tail-sampling-strategy}.

\bibitem{newrelicTailbasedLimits}
{New Relic}.
\newblock {Technical distributed tracing details: Trace limits}.
\newblock Retrieved April 2022 from
  \url{https://docs.newrelic.com/docs/distributed-tracing/concepts/how-new-relic-distributed-tracing-works/#limits}.

\bibitem{newrelicTailbased}
{New Relic}.
\newblock {Tail-based sampling (Infinite Tracing)}.
\newblock Retrieved April 2022 from
  \url{https://docs.newrelic.com/docs/understand-dependencies/distributed-tracing/get-started/how-new-relic-distributed-tracing-works#tail-based},
  2020.

\bibitem{opentelemetry}
{OpenTelemetry: An Observability Framework for Cloud-Native Software}.
\newblock Retrieved April 2022 from \url{http://opentelemetry.io/}.

\bibitem{otTailSampling}
{OpenTelemetry}.
\newblock {Tail Sampling Processor}.
\newblock Retrieved April 2022 from
  \url{https://github.com/open-telemetry/opentelemetry-collector-contrib/tree/main/processor/tailsamplingprocessor}.

\bibitem{ots307}
{OpenTelemetry Specification Issue 307: Allow samplers to be called during
  different moments in the Span lifetime}.
\newblock Retrieved April 2022 from
  \url{https://github.com/open-telemetry/opentelemetry-specification/issues/307},
  2019.

\bibitem{oteps115}
{OpenTelemetry Enhancement Proposal 115: Allow Additional Sampling Hooks}.
\newblock Retrieved April 2022 from
  \url{https://github.com/open-telemetry/oteps/pull/115}, 2020.

\bibitem{opentracing}
{OpenTracing: Vendor-Neutral APIs and Instrumentation for Distributed Tracing}.
\newblock Retrieved April 2022 from \url{http://opentracing.io/}.

\bibitem{ostrowski2011diagnosing}
Krzysztof Ostrowski, Gideon Mann, and Mark Sandler.
\newblock Diagnosing latency in multi-tier black-box services.
\newblock In {\em 4th International Workshop on Large-Scale Distributed Systems
  and Middleware (LADIS'11)}, 2011.

\bibitem{netflixTracing}
Maulik Pandey.
\newblock {Building Netflix’s Distributed Tracing Infrastructure}.
\newblock Retrieved April 2022 from
  \url{https://netflixtechblog.com/building-netflixs-distributed-tracing-infrastructure-bb856c319304},
  2019.

\bibitem{parker2020distributed}
Austin Parker, Daniel Spoonhower, Jonathan Mace, Ben Sigelman, and Rebecca
  Isaacs.
\newblock {\em Distributed Tracing in Practice: Instrumenting, Analyzing, and
  Debugging Microservices}.
\newblock O'Reilly Media, 2020.

\bibitem{sambasivan2016principled}
Raja~R Sambasivan, Ilari Shafer, Jonathan Mace, Benjamin~H Sigelman, Rodrigo
  Fonseca, and Gregory~R Ganger.
\newblock Principled workflow-centric tracing of distributed systems.
\newblock In {\em Proceedings of the Seventh ACM Symposium on Cloud Computing},
  pages 401--414, 2016.

\bibitem{sambasivan2011diagnosing}
Raja~R Sambasivan, Alice~X Zheng, Michael De~Rosa, Elie Krevat, Spencer
  Whitman, Michael Stroucken, William Wang, Lianghong Xu, and Gregory~R Ganger.
\newblock Diagnosing performance changes by comparing request flows.
\newblock In {\em 8th USENIX Symposium on Networked Systems Design \&
  Implementation (NSDI'11)}, volume~5, pages 1--1, 2011.

\bibitem{shkuro2019mastering}
Yuri Shkuro.
\newblock {\em Mastering Distributed Tracing}.
\newblock Packt Publishing, Feb 2019.

\bibitem{hdfs}
Konstantin Shvachko, Hairong Kuang, Sanjay Radia, and Robert Chansler.
\newblock The {Hadoop} distributed file system.
\newblock In {\em 2010 IEEE 26th symposium on mass storage systems and
  technologies (MSST)}, pages 1--10. Ieee, 2010.

\bibitem{sigelman2010dapper}
Benjamin~H. Sigelman, Luiz~André Barroso, Mike Burrows, Pat Stephenson, Manoj
  Plakal, Donald Beaver, Saul Jaspan, and Chandan Shanbhag.
\newblock Dapper, a large-scale distributed systems tracing infrastructure.
\newblock Technical report, Google, Inc., 2010.

\bibitem{splunkTraceLimits}
{Splunk}.
\newblock {Observability Cloud Usage, Subscription Limits Enforcement, and
  Entitlements}.
\newblock Retrieved April 2022 from
  \url{https://www.splunk.com/en_us/legal/usage-subscription-limits-enforcement-and-entitlements.html},
  2022.

\bibitem{splunkLoggingPerformance}
{Splunk}.
\newblock {Splunk Enterprise Capacity Planning Manual - Summary of performance
  recommendations}.
\newblock Retrieved April 2022 from
  \url{https://docs.splunk.com/Documentation/Splunk/8.2.6/Capacity/Summaryofperformancerecommendations},
  2022.

\bibitem{splunkUseCases}
{Splunk}.
\newblock {Use cases: Troubleshoot errors and monitor application performance
  using Splunk APM}.
\newblock Retrieved April 2022 from
  \url{https://docs.splunk.com/Observability/apm/apm-use-cases/apm-use-cases-intro.html#nav-Use-cases:-Troubleshoot-errors-and-monitor-application-performance},
  2022.

\bibitem{sridharan2018distributed}
Cindy Sridharan.
\newblock {\em Distributed Systems Observability}.
\newblock O'Reilly Media, 2018.

\bibitem{suo2016time}
Kun Suo, Jia Rao, Luwei Cheng, and Francis~CM Lau.
\newblock Time capsule: Tracing packet latency across different layers in
  virtualized systems.
\newblock In {\em Proceedings of the 7th ACM SIGOPS Asia-Pacific Workshop on
  Systems}, pages 1--9, 2016.

\bibitem{thereska2006stardust}
Eno Thereska, Brandon Salmon, John Strunk, Matthew Wachs, Michael Abd-El-Malek,
  Julio Lopez, and Gregory~R Ganger.
\newblock {Stardust: Tracking Activity in a Distributed Storage System}.
\newblock In {\em 2006 ACM International Conference on Measurement and Modeling
  of Computer Systems (SIGMETRICS '06)}, 2006.

\bibitem{verma2015borg}
Abhishek Verma, Luis Pedrosa, Madhukar~R. Korupolu, David Oppenheimer, Eric
  Tune, and John Wilkes.
\newblock Large-scale cluster management at {Google with Borg}.
\newblock In {\em Proceedings of the 10th European Conference on Computer
  Systems (EuroSys'15)}, Bordeaux, France, 2015.

\bibitem{wu2019zeno}
Yang Wu, Ang Chen, and Linh Thi~Xuan Phan.
\newblock Zeno: diagnosing performance problems with temporal provenance.
\newblock In {\em 16th USENIX Symposium on Networked Systems Design and
  Implementation (NSDI'19)}, pages 395--420, 2019.

\bibitem{yin2011empirical}
Zuoning Yin, Xiao Ma, Jing Zheng, Yuanyuan Zhou, Lakshmi~N Bairavasundaram, and
  Shankar Pasupathy.
\newblock An empirical study on configuration errors in commercial and open
  source systems.
\newblock In {\em Proceedings of the 23rd ACM Symposium on Operating Systems
  Principles (SOSP'11)}, pages 159--172, 2011.

\bibitem{zhang2016treadmill}
Yunqi Zhang, David Meisner, Jason Mars, and Lingjia Tang.
\newblock Treadmill: Attributing the source of tail latency through precise
  load testing and statistical inference.
\newblock In {\em 2016 ACM/IEEE 43rd Annual International Symposium on Computer
  Architecture (ISCA)}, pages 456--468. IEEE, 2016.

\bibitem{zhou2011secure}
Wenchao Zhou, Qiong Fei, Arjun Narayan, Andreas Haeberlen, Boon~Thau Loo, and
  Micah Sherr.
\newblock Secure network provenance.
\newblock In {\em Proceedings of the twenty-third ACM symposium on operating
  systems principles}, pages 295--310, 2011.

\bibitem{zhou2012distributed}
Wenchao Zhou, Suyog Mapara, Yiqing Ren, Yang Li, Andreas Haeberlen, Zachary
  Ives, Boon~Thau Loo, and Micah Sherr.
\newblock Distributed time-aware provenance.
\newblock {\em Proceedings of the VLDB Endowment}, 6(2):49--60, 2012.

\bibitem{zhou2010efficient}
Wenchao Zhou, Micah Sherr, Tao Tao, Xiaozhou Li, Boon~Thau Loo, and Yun Mao.
\newblock Efficient querying and maintenance of network provenance at
  internet-scale.
\newblock In {\em Proceedings of the 2010 ACM SIGMOD International Conference
  on Management of data}, pages 615--626, 2010.

\bibitem{zipkin}
{Zipkin: A Distributed Tracing System}.
\newblock Retrieved April 2022 from \url{http://zipkin.io/}.

\end{thebibliography}

\end{document}